\documentclass[sigconf]{acmart}

\usepackage[utf8]{inputenc} 
\usepackage[T1]{fontenc}    
\usepackage{hyperref}       
\usepackage{url}            
 
\usepackage{booktabs}       
\usepackage{amsfonts}       
\usepackage{nicefrac}       
\usepackage{microtype}      
\usepackage{xcolor}         
\usepackage{multirow}
\usepackage{enumitem}
\usepackage{xcolor}
\usepackage{booktabs}
\usepackage{graphicx}
\usepackage{textcomp}
\usepackage{multirow}
\usepackage{xcolor}
\usepackage{caption}
\usepackage{color}
\usepackage{float} 
 \usepackage{hyperref}

\usepackage{algorithm}  
\usepackage{algpseudocode}  
\usepackage{amsmath}  

\AtBeginDocument{%
  \providecommand\BibTeX{{%
    \normalfont B\kern-0.5em{\scshape i\kern-0.25em b}\kern-0.8em\TeX}}}

\copyrightyear{2023}
\acmYear{2023}
\setcopyright{acmlicensed}\acmConference[KDD '23]{Proceedings of the 29th
ACM SIGKDD Conference on Knowledge Discovery and Data Mining}{August
6--10, 2023}{Long Beach, CA, USA}
\acmBooktitle{Proceedings of the 29th ACM SIGKDD Conference on Knowledge
Discovery and Data Mining (KDD '23), August 6--10, 2023, Long Beach, CA,
USA}
\acmPrice{15.00}
\acmDOI{10.1145/3580305.3599789}
\acmISBN{979-8-4007-0103-0/23/08}

\title{CBLab: Supporting the Training of Large-scale Traffic Control Policies with Scalable Traffic Simulation}



\begin{document}

\title{CBLab: Supporting the Training of Large-scale Traffic Control Policies with Scalable Traffic Simulation}


\author{Chumeng Liang}
\email{caradryan2022@gmail.com}
\affiliation{%
 \institution{Shanghai Jiao Tong University}
 \country{China}
}

\author{Zherui Huang}
\email{huangzherui@sjtu.edu.cn}
\affiliation{%
 \institution{Shanghai Jiao Tong University}
 \country{China}
}

\author{Yicheng Liu}
\email{liuyicheng1515@sjtu.edu.cn}
\affiliation{%
 \institution{Shanghai Jiao Tong University}
 \country{China}
}

\author{Zhanyu Liu}
\email{zhyliu00@sjtu.edu.cn}
\affiliation{%
 \institution{Shanghai Jiao Tong University}
 \country{China}
}

\author{Guanjie Zheng}
\authornote{corresponding author}
\email{gjzheng@sjtu.edu.cn}
\affiliation{%
 \institution{Shanghai Jiao Tong University}
 \country{China}
}

\author{Hanyuan Shi}
\email{shihanyuan@sjtu.edu.cn}
\affiliation{%
 \institution{Independent Researchers}
 \country{China}
}

\author{Kan Wu}
\email{kanwu@zhejianglab.com}
\affiliation{%
 \institution{Research Center for Intelligent Transportation, Zhejiang Lab}
 \country{China}
}

\author{Yuhao Du}
\email{apiadu17a6@gmail.com}
\affiliation{%
 \institution{Independent Researchers}
 \country{China}
}

\author{Fuliang Li}
\email{tjfulianglee@gmail.com}
\affiliation{%
 \institution{Baidu}
 \country{China}
}

\author{Zhenhui Li}
\email{jessielzh@gmail.com}
\affiliation{%
 \institution{Yunqi Academy of Engineering}
 \country{China}
}

\renewcommand{\shortauthors}{Liang et al.}

\begin{abstract}
  Traffic simulation provides interactive data for the optimization of traffic control policies. However, existing traffic simulators are limited by their lack of scalability and shortage in input data, which prevents them from generating interactive data from traffic simulation in the scenarios of real large-scale city road networks. 

In this paper, we present \textbf{C}ity \textbf{B}rain \textbf{Lab}, a toolkit for scalable traffic simulation. CBLab consists of three components: CBEngine, CBData, and CBScenario. CBEngine is a highly efficient simulator supporting large-scale traffic simulation. CBData includes a traffic dataset with road network data of 100 cities all around the world. We also develop a pipeline to conduct a one-click transformation from raw road networks to input data of our traffic simulation. Combining CBEngine and CBData allows researchers to run scalable traffic simulations in the road network of real large-scale cities. Based on that, CBScenario implements an interactive environment and a benchmark for two scenarios of traffic control policies respectively, with which traffic control policies adaptable for large-scale urban traffic can be trained and tuned. To the best of our knowledge, CBLab is the first infrastructure supporting traffic control policy optimization in large-scale urban scenarios. CBLab has supported the City Brain Challenge @ KDD CUP 2021. The project is available on GitHub:~\url{https://github.com/CityBrainLab/CityBrainLab.git}.
\end{abstract}



\maketitle

\section{Introduction}
\label{sec:motivation}
\begin{figure*}
\centering
  \includegraphics[width=0.95\linewidth]{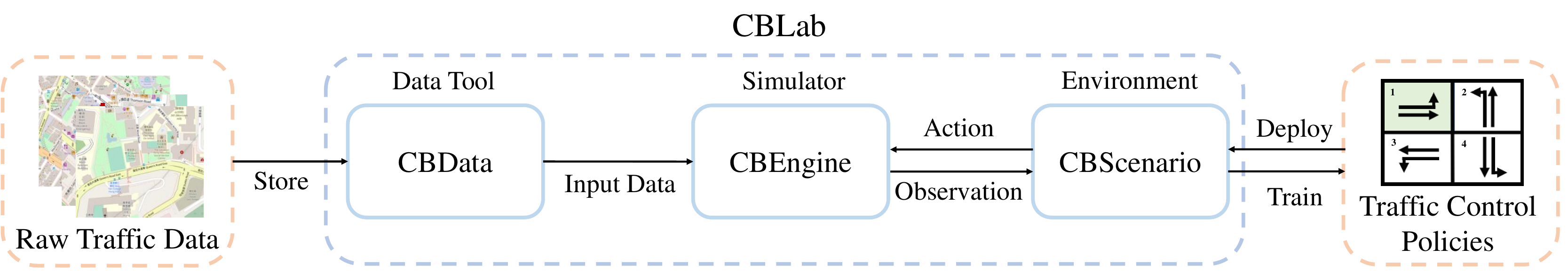}
  \caption{An overview of CBLab.}
  \label{fig:overview}
  \vspace{-0.3cm}
\end{figure*}

Well-crafted traffic control policies, such as traffic signal control and congestion pricing, are expected to improve the efficiency of urban transportation. In recent years, many studies have been conducted to optimize the traffic control policies according to real-time traffic data~\cite{haarnoja2018soft,wei2018intellilight,wei2019presslight,zheng2019learning,zhang2020generalight,chen2020toward,zang2020metalight,oroojlooy2020attendlight,wu2021dynstgat, zhang2022expression,edtoll,EBGtoll}. These policies depend on data generated by interaction with the traffic environment where they explore to make good decisions under different consequences.

However, real-world urban traffic cannot provide enough interactive data to train these policies, because the exploration of the policy may have a toxic impact on the urban traffic e.g. provoke severe congestion. Traffic simulators are therefore born as alternatives to provide traffic environments for traffic control policies to interact with. These simulators~\cite{SUMO2018,zhang2019cityflow,qarsumo} simulate the microscopic evolution of the urban traffic. For each time step, they describe the traffic state, obtain a traffic action from the decision made by traffic control policies and make it happen in the simulation. Traffic control policies can then learn from how the traffic evolves under certain actions and improve decision-making.

While existing traffic simulators help hatch various traffic control policies successfully, they still come with drawbacks. Current simulators, as they were designed primitively, support simulation in road networks smaller than one hundred intersections and cannot scale to city-level traffic, which involves thousands of intersections. Due to limits in efficiency and scalability, these simulators are either not able to conduct a city-level simulation in a feasible time or set to prevent masses of vehicles from coming in the traffic. 

\begin{figure*}[htbp]
			\centering
	\includegraphics[width=0.95\linewidth]{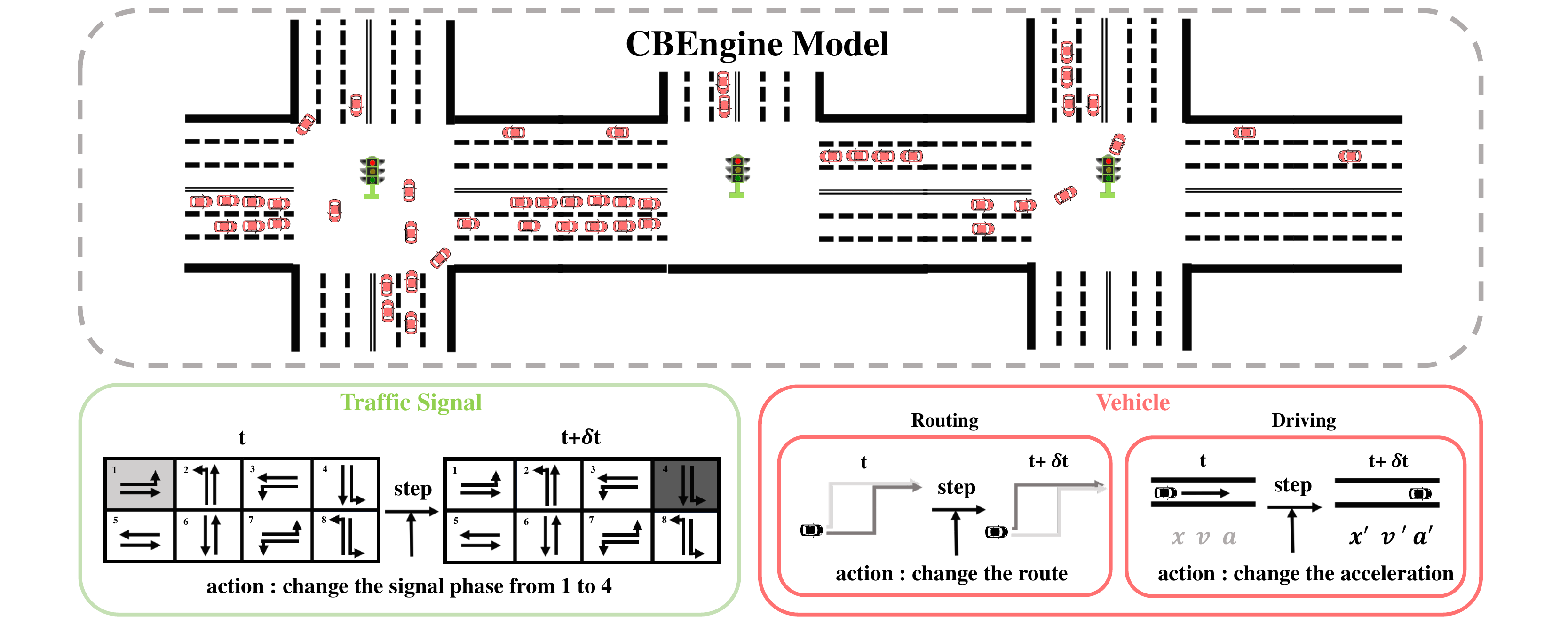}
	\centering
	\caption{The Traffic Model of CBEngine}
	\label{fig:simulator}
 \vspace{-0.3cm}
\end{figure*}

Another concern lies in the shortage of input data for large-scale traffic simulation. Although the map data of the main cities in the world is completed, there is an absence of infrastructure for access to the map data and a pipeline to transform it into simulation inputs. Therefore, inputs for traffic simulation only come from manual work and are limited to a small set of road networks~\cite{wei2019survey,wei2019colight,zheng2019learning,xu2021hierarchically} whose scales are often dozens of intersections (e.g. 4x3 or 4x4) - much smaller than real urban road networks. 

To overcome two aforementioned drawbacks, we propose \textbf{C}ity \textbf{B}rain \textbf{Lab}, a novel toolkit for scalable traffic simulation. CBLab consists of three components: a microscopic traffic simulator CBEngine, a data tool CBData, and a traffic control policy environment CBScenario (See Figure~\ref{fig:overview}). Benefiting from well-designed parallelization, CBEngine is of high efficiency and scalability and is capable of running the traffic simulation on the scale of 10,000 intersections and 100,000 vehicles with a 
real-simulation time ratio of 1:4 with ordinary computing hardware. CBData includes an accessible dataset that contains raw road networks of 100 main cities all around the world. A pipeline is prepared to automatically transform the raw data into input data for traffic simulation. Combining CBEngine and CBData, users can easily start up traffic simulation on real city-level road networks. Based on the scalable traffic simulation, we implement CBScenario as an environment for two traffic control policies: traffic signal control and congestion pricing. Users can design, develop, and train traffic control policies in the framework of CBScenario. To the best of our knowledge, we are the first to provide infrastructure for large-scale traffic control policy optimization.

Our contribution can be summarized as follows.
\begin{itemize}
\item We develop a scalable traffic simulator CBEngine that supports city-level microscopic traffic simulation for the first time.
\item We develop a data tool CBData to provide input data for large-scale traffic simulation. 
\item We implement an interactive environment CBScenario for training traffic control policies under a large-scale setting.
\end{itemize}

The original version of CBLab has successfully supported the City Brain Challenge @ KDD CUP 2021 with 1,156 participating teams. See Appendix~\ref{appendix:kddcup} for details.

\section{CBEngine: City-Scale Traffic Simulation Engine}
\label{sec:simulator}

In this section, we introduce CBEngine, the traffic simulator. We demonstrate its traffic modeling and conduct extensive experiments to show the efficiency, scalability, and plausibility of CBEngine.  

\subsection{Overview of Simulation Modeling}
\label{sec:cbenginemodel}
Traffic simulators take the road network and the traffic flow (vehicles in the traffic) as inputs and aim to simulate their interaction. Road networks describe the topology of roads and intersections. Traffic flows describe the origins, destinations, and routes of vehicles. As shown in Figure~\ref{fig:simulator}, the road network interacts with vehicles through traffic signal lights, which control the passing of vehicles at intersections. When the simulation starts, vehicles in the traffic flow set out from their origins, travel down the routes, and finally arrive at their destinations. Roads, intersections, traffic signal lights, and vehicles can be considered as \textit{traffic elements}.

We can formulate the simulation in the form of states and actions. The simulator holds states for traffic elements at the time step $t$. For the vehicle, the state is its location and speed. For the traffic signal light, the state is the current signal. Actions of elements then iterate the current state to that in the next time step. For example, the acceleration of a vehicle rises up the vehicle's speed at the next time step. Finally, the simulation moves on to the next time step, where a new state-action iteration will begin. Let $s_t$, $a_t$, and $s_{t+1}$ denote the current state, action, and the next state, the traffic modeling in the simulation can be concluded by Eq~\ref{only}.

\begin{equation}
\label{only}
 s_{t} + a_{t} \rightarrow s_{t+1}
\end{equation}

\subsection{Road Network} 
In CBEngine, \textbf{Road Network} is the topological network where vehicles drive. It consists of two components: roads and intersections. \textbf{Road} models the road segment in the real-world road network. A road may include multiple lanes. Each \textbf{lane} holds one or more vehicles. \textbf{Intersection} is the nexus of different roads. Through \textbf{lane links} in the intersection, lanes of different roads connect to each other. \textbf{Traffic signal light} is another key element in the intersection, which assigns a true-or-false signal for each lane link. Vehicles can only move from one road to another through available lane links with a true signal. 

\subsection{Driving Model}

Behaviors of vehicles are controlled by the driving model in the traffic simulation. Driving models determine how vehicles move on the road. Towards simulating the behaviors of vehicles, researchers have proposed various models~\cite{krauss1998microscopic,yuan2010t,yuan2011driving}.

The default driving model of CBEngine is a modified version of the driving model used in Cityflow~\cite{zhang2019cityflow}, originating from the driving model proposed by Stefan Krauß~\cite{krauss1998microscopic}. The key idea is that: the vehicle will drive as fast as possible subject to safety regularization. Specifically, the maximum speed of the vehicle is subject to several static or dynamic speed constraints. Vehicles will accelerate or decelerate to the speed with the maximum acceleration or deceleration. In the implementation, the maximum acceleration (deceleration) is a hyperparameter and is open to users to tune with an API. The considered speed constraints are listed and discussed respectively as follows:
\begin{itemize}
    \item Road speed limit
    \item Collision-free following \& leading speed
    \item Cutting-in collision-free speed
    \item Traffic-signal safe speed
\end{itemize}
\vspace{-0.2cm}
\paragraph{Road speed limit} Each road has its own speed limit. This is a static speed constraint.
\vspace{-0.2cm}
\paragraph{Collision-free following \& leading speed} To avoid collisions, vehicles need to adapt their speed to the speed of their following and leading vehicles. We use the collision-free following speed to model the max speed constrained by the leading vehicle. Following the driving model in Cityflow, we compute these two constraints with Eq~\ref{eq:cfs}. It takes $v$ current speed of the vehicle, $v_L$ current speed of the leading vehicle, $d$ maximum deceleration of the vehicle, $d_L$ maximum deceleration of the leading vehicle, $D$ the current distance between two vehicles, $interval$ the length of each time step as parameters to compute the collision-free following speed $s_{cfs}$. 
\begin{equation}
\begin{aligned}
\label{eq:cfs}
    a &= \frac{1}{2\cdot d}, b= \frac{interval}{2}\\
    c &= \frac{v\cdot interval}{2}-\frac{v_L^2}{2\cdot d_L} - D\\
    s_{cfs} &= \frac{-b + \sqrt{b^2-4ac}}{2\cdot a}
\end{aligned}
\end{equation}

To compute the collision-free leading speed $s_{cls}$, we use a mirror equation of Eq~\ref{eq:cfs}. We use the $d_F$ the maximum deceleration of the following vehicle and $v_F$ the speed of the following vehicle to replace $d_L$ and $v_L$ in the Eq~\ref{eq:cfs}. The collision-free following and leading speed constraints can be summarized as Eq~\ref{eq:cs}.

\begin{equation}
\begin{aligned}
\label{eq:cs}
    s_{cls} \leq v\leq s_{cfs}
\end{aligned}
\end{equation}

\paragraph{Cutting-in collision-free speed} CBEngine supports self-adaptive lane changing within the road (Cutting-in). This asks for cutting-in collision-free following and leading speed, avoiding collisions with the leading and following vehicles in the target lane. We compute this constraint using Eq~\ref{eq:cfs} with $v_L$, $d_L$, and $D$ given by the leading vehicle in the target lane and $v_F$, $d_F$, and $D$ given by the following vehicle in the target lane. 

An exception happens when the vehicle needs to conduct an emergent cutting-in. This takes place when the vehicle is very close to the intersection but still in the wrong lane (e.g. The vehicle is in the go-straight lane but needs to turn left). On this occasion, the vehicle tries to stop to wait until it is able to change the lane. Therefore, the maximum speed constraint equals zero and the vehicle will decelerate to zero with its maximum deceleration. 
\vspace{-0.2cm}
\paragraph{Traffic-signal safe speed} 
Vehicles heading for an intersection are subject to two constraints determined by the traffic signal. First, the current speed can be decelerated to zero within the remained passing time of the traffic signal. Second, the driving distance cannot exceed the distance to the intersection, under the decelerating process defined in the first constraint.



\begin{figure*}[t]
			\centering
			\includegraphics[width=0.9\linewidth]{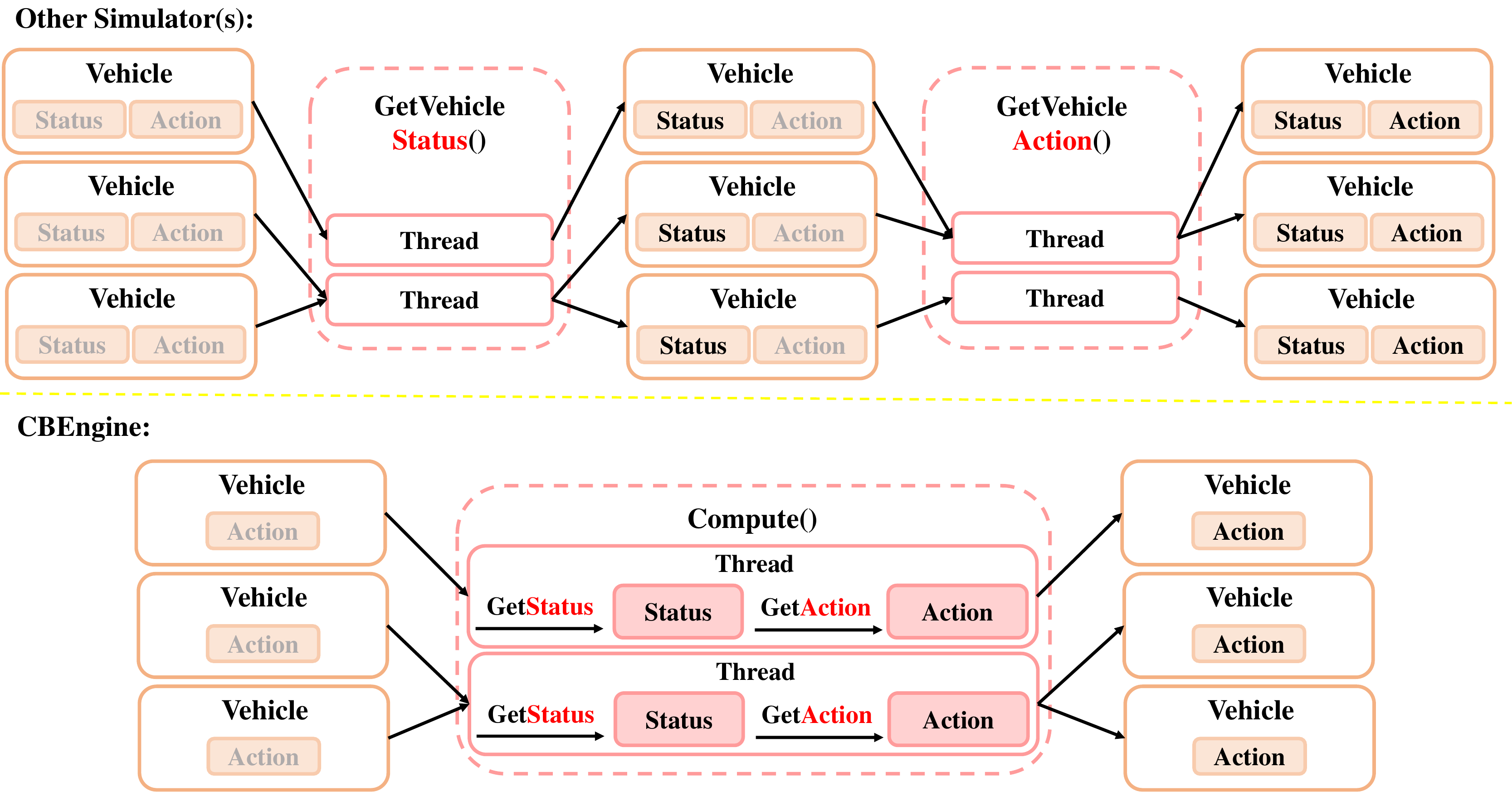}
	\centering
	\caption{Comparison between CBEngine and other simulators on parallelization design of computing actions for vehicles. The gray data member indicates it is not updated, while the black one is updated. }
	\label{fig:app}
 \vspace{-0.3cm}
\end{figure*}

\begin{figure*}[t]
			\centering
			\includegraphics[width=\linewidth]{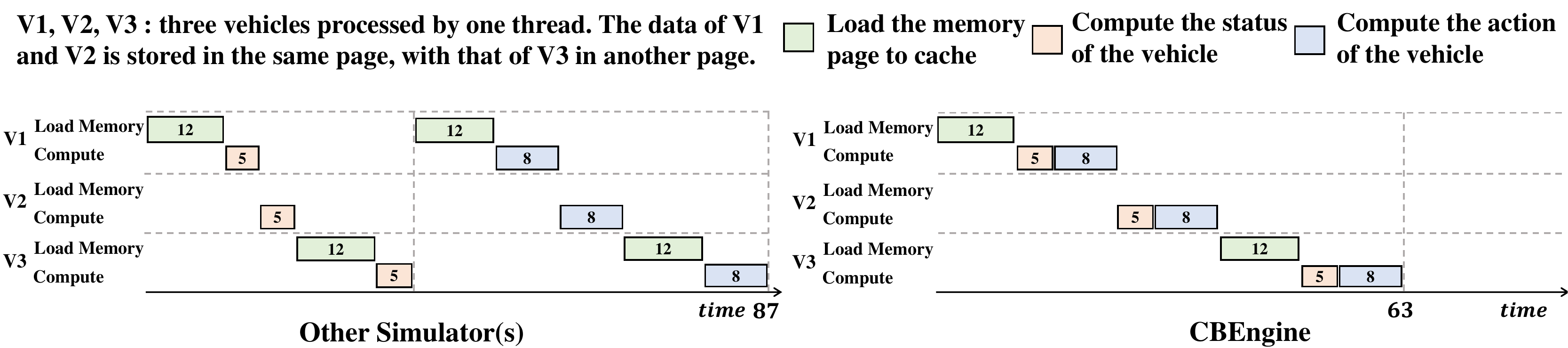}
	\centering
	\caption{A schedule case of computing vehicle action. The length of the bar indicates the time the corresponding task consumes.}
	\label{fig:sche}
 \vspace{-0.5cm}
\end{figure*}

\subsection{Model Customization}
An important function implemented in CBEngine is vehicle model customization. In the simulator, both the driving model and the routing model are black boxes. We modularize two models as an independent C++ class in our implementation, respectively. Users can customize two models by modifying the class without modification to other parts of CBEngine. Our documentation in Appendix~\ref{appendix:doc} provides detailed instructions. To the best of our knowledge, CBEngine is the first traffic simulator to support easy-to-use customization for driving and routing models.     

\subsection{Efficiency and Scalability}
\label{subsec:eff}
The major bottleneck of supporting city-wide traffic control policy training is the limit in efficiency and scalability of existing simulators, of which extensive efforts have been made in CBEngine for improvement. The progress in efficiency and scalability can be validated by the time cost of running the simulation on the same traffic and the maximum capacity of intersections and vehicles. In this subsection, we first introduce several systematic designs of CBEngine that greatly promote its efficiency and scalability. Then, we conduct extensive experiments to validate the efficiency and scalability of CBEngine. We mainly compare CBEngine with two widely-used traffic simulators, SUMO~\cite{SUMO2018} and Cityflow~\cite{zhang2019cityflow}.

\subsubsection{System Implementation}
To overcome the disadvantages in efficiency and scalability, we introduce three designs for the system implementation of CBEngine, among which \textit{Parallel Design in Computing the Vehicle Behavior} mainly contributes to the efficiency while \textit{Lane Changing in Driving Model} and \textit{Intersection Links} contribute to the scalability.

\begin{table*}[thbp]
\label{tab:efficiency}
\begin{tabular}{c|c|c|c|c|c|c}
\toprule
Dataset   & Nanchang & Changchun & JiNan & Shenzhen & Hangzhou & Shanghai\\ \midrule
Intersection Num  & 1506 & 2228 & 2314 & 3427 & 3434 & 8474\\
\midrule
Vehicle Num   & \textasciitilde 25000 & \textasciitilde 50000 & \textasciitilde 50000 & \textasciitilde 70000 & \textasciitilde 70000 &  \textasciitilde 120000\\ \midrule
\midrule
 & \multicolumn{6}{c}{Time Cost} \\ \midrule
\multirow{2}{*}{SUMO} & 1239.93 & 2091.60 & 2151.01 & 3103.58 & 3199.14 & 6173.51 \\  
& ($\pm3.58$) & ($\pm$7.84) & ($\pm70.64$) & ($\pm110.08$) & ($\pm70.87$) & ($\pm75.27$) \\
\midrule
\multirow{2}{*}{Cityflow} & 164.08 & 243.22 & 242.47 & 289.31 & 310.98 & 664.18 \\
  & ($\pm$6.30) & ($\pm$11.85) & ($\pm$21.34) & ($\pm$\textbf{1.74}) & ($\pm$18.22) & ($\pm$21.01) \\
\midrule
\midrule
\multirow{2}{*}{CBEngine} & \textbf{104.39} & \textbf{169.36} & \textbf{174.29} & \textbf{243.26} & \textbf{245.01} & \textbf{472.07} \\
& ($\pm$\textbf{0.54}) & ($\pm$\textbf{1.60}) & ($\pm$\textbf{1.93}) & ($\pm$1.75) & ($\pm$\textbf{2.16}) & ($\pm$\textbf{2.90}) \\
\bottomrule

\end{tabular}
\caption{Comparison results of efficiency experiments in six real-world cities.}
\label{tab:efficiency}
\vspace{-0.3cm}
\end{table*}

\paragraph{Parallel Design in Computing the Vehicle Behavior}
The future state in the simulation is determined by the present state and the current action. This is the major computational job in traffic simulation, which costs over $90\%$ of the step time. In CBEngine, we carefully design the parallelization process of computing the action for vehicles. The comparison of our design and that of other simulators is shown in Figure~\ref{fig:app}.

The process to compute actions for vehicles can be divided into two stages: getting the Status and getting the Action. In the first stage, the simulator collects information for computing the action of a vehicle. The set of information is denoted as the \textit{Status}. In the second stage, the simulator computes the action according to Status. Existing simulators parallelize these two stages respectively. Vehicle objects in the simulator need to keep two data members: the Status and the Action. Two members are updated sequentially by two parallelized methods: GetVehicleStatus() and GetVehicleAction().

In CBEngine, we assemble two stages in one parallelized method: Compute(). We implement this by adjusting data dependencies and reconstructing the parallel architecture. This design benefits efficiency from two perspectives. First, the space cost is lower. We bind the Status data on the Thread object rather than on the Vehicle objects. This is because the number of threads is quite smaller than that of vehicles. Second, by merging two stages, the CPU is more likely to access existing pages in the memory when getting the action, because those pages are recently loaded to the memory when getting the status. Since this design reduces the number of times that the CPU loads pages from the memory, it decreases the number of cache misses and the time cost. 

Figure~\ref{fig:sche} raises a case comparing the scheduling of one thread processing vehicles in other simulators to that in CBEngine. For each vehicle, the thread needs to compute its status and then its action accordingly. The page where the vehicle is stored is required to be loaded in memory for access to the vehicle. Assume that the cache can only store one page. For other simulators, the processing is divided into two stages, while each stage loads two pages from memory. This is because the thread needs to access all three vehicles in each stage. By contrast, processing in CBEngine combines two stages and does not access new vehicles until the job on the vehicle is finished. This design helps reduce the operation of loading pages and saves processing time.

\paragraph{Lane Changing in Driving Model} \label{sec:lanechanging} Lane changing is a driving action. Drivers may change the lane if the current lane is too congested. However, lane changing is hard to simulate. Cityflow~\cite{zhang2019cityflow}, one of our baselines, omits all lane changing except those happening at the intersection. The lane of the vehicle is determined by the direction it will turn. This simplifies the implementation but leads to poor plausibility because vehicles on the road cannot make full use of all lanes as they do in real urban traffic. They have to stay at the current lane and may wait a long time, although their neighbor lane is clear. Furthermore, to avoid collision on this occasion, Cityflow does not allow new vehicles to come in until the lane is relatively unblocked. This limitation severely impacts the scalability of Cityflow and explains the fact in our experiment of scalability that Cityflow cannot hold 1,000,000 vehicles.

In CBEngine, we implement a driving model allowing lane changing. Vehicles are put in a random lane when they get into the road. They will try lane changing according to the direction they are to turn to. But if that lane is in congestion, they will keep going in the current lane which is relatively clear. This design achieves higher plausibility and provides stronger scalability for CBEngine.

\paragraph{Intersection Links} Another key mechanism of the traffic simulator is the intersection link. SUMO and Cityflow track the behavior of vehicles inside the intersection. However, due to the limit in the implementation, this design may lead to deadlocks very frequently in the practice, especially when running large-scale traffic simulations. This is because the track of some vehicles may block that of others. Therefore, we conduct simplification here to avoid such deadlocks. When a vehicle passes the intersection, the intersection will hold it for a while and then send it to the target road. We believe that the effect of the intersection link can be simulated by the holding time. To the best of our knowledge, our design avoids all such deadlocks in practice.

\subsubsection{Experiments}
\label{sec:effiexp}
\begin{figure}[tbp]
			\centering
			\includegraphics[width=0.85\linewidth]{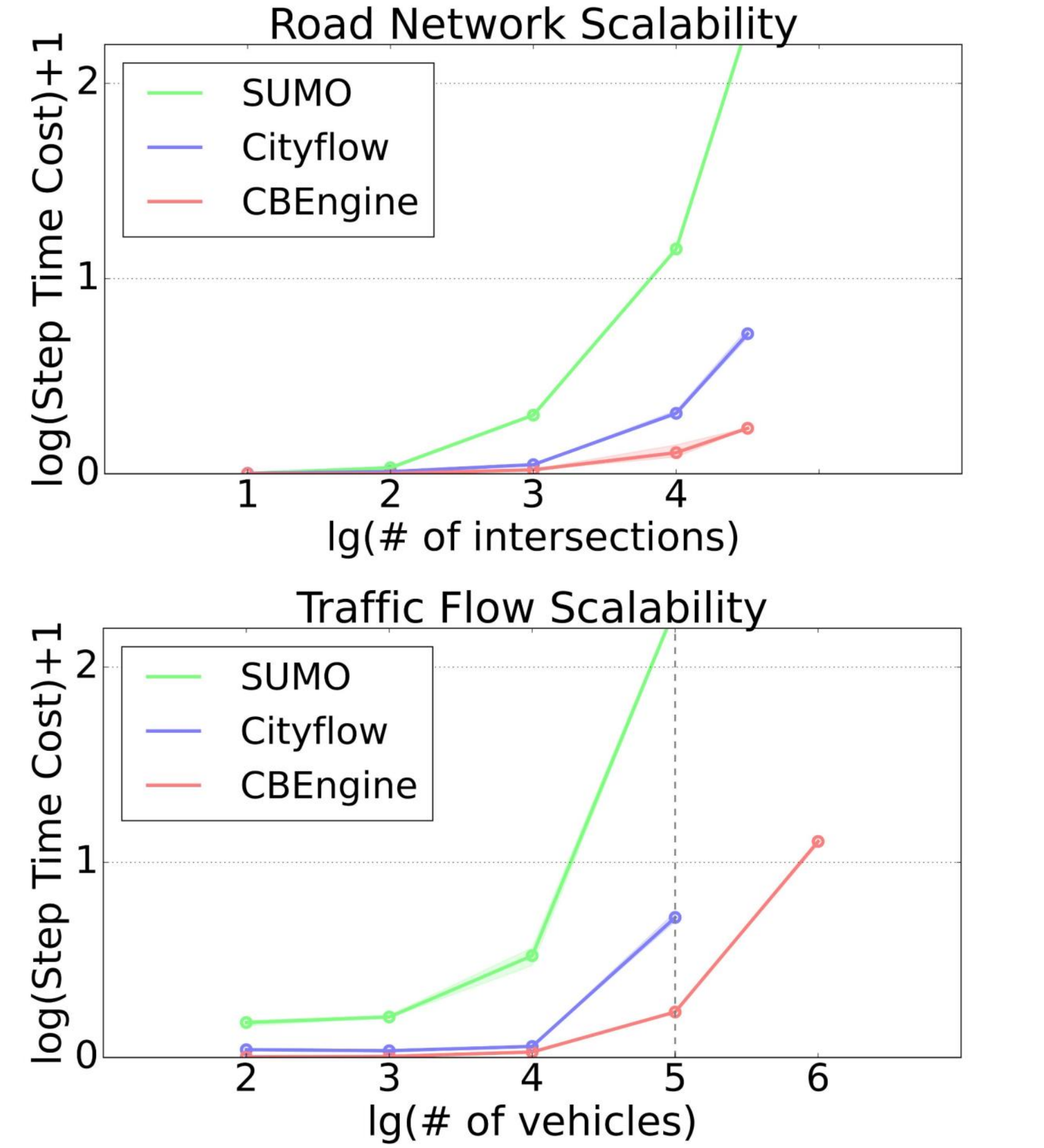}
	\centering
	\caption{Comparison results of experiments for scalability.}
	\label{fig:scalability}
 \vspace{-0.6cm}
\end{figure}

\paragraph{Experimental Setup}
To evaluate the efficiency and scalability of CBEngine, we compare it with two widely-used open-source microscopic traffic simulators, SUMO~\cite{SUMO2018} and Cityflow~\cite{zhang2019cityflow}. We compare these simulators in three aspects, running time, road network scalability, and traffic flow scalability. All the experiments are conducted on a Ubuntu20.04 system with a 40-core CPU and 128GB RAM. The unit of time cost is second for all three experiments. More details of the experimental setup are given in Appendix~\ref{appendix:exp}.
\label{checklist:setup}

\paragraph{Experiment 1: Efficiency}
We run a one-hour traffic simulation on urban traffic cases of six cities with distinct scales and compare the time cost of baselines and our simulator. Road networks of these cases are obtained and cleaned from OpenStreetMap~\cite{haklay2008openstreetmap}. Traffic flows are generated according to the scale of the road network. Results are displayed in Table~\ref{tab:efficiency}.   

We can observe that CBEngine achieves significant improvement in simulation efficiency (usually 30\%-40\% compared with Cityflow and more than 90\% compared with SUMO). The stability of CBEngine is distinctly better than that of baselines. Furthermore, the gap between baselines and ours grows with the scale of traffic cases, indicating that CBEngine can adapt well to large-scale cases.

\paragraph{Experiment 2: Scalability on Road Networks}
Simulators with high scalability on road networks can run the simulation efficiently on large-scale road networks. To explore the scalability of baselines and CBEngine on road networks, we run a traffic simulation on road networks of different scales. We select five regions from real road networks with $\{10, 10^{2}, 10^{3}, 10^{4}, 10^{4.5}\}$ intersections. The upper bound is set as $10^{4.5}$ because this is the largest road network for a single city in OpenStreetMap. For each setting, we repeat the experiment three times. We report the time cost of single-step simulation for baselines and CBEngine as well as the range. 

The results are visualized in Figure~\ref{fig:scalability} (left). CBEngine outperforms two baselines in time cost on road networks with all scales. Take the experiment setup under the road network with the largest scale as a quantitative example. The average single step time cost of CBEngine is 0.2670 seconds, while that of SUMO~\cite{SUMO2018} and Cityflow~\cite{zhang2019cityflow} are 9.1832 seconds and 1.0343 seconds, respectively.
\paragraph{Experiment 3: Scalability on Traffic Flows}
With high scalability on traffic flows, the simulator keeps efficient under heavy traffic. Similar to Experiment 2, we conduct an experiment to evaluate the scalability of traffic flows of baselines and CBEngine. We generate five traffic flows with $\{10^{2}, 10^{3}, 10^{4}, 10^{5}, 10^{6}\}$ on-way vehicles. For each setting, we repeat the experiment three times. We report the time cost of a single step as well as the range.

The results are visualized in Figure~\ref{fig:scalability} (right). CBEngine outperforms two baselines under different scales of traffic flows. To give a quantitative example, the average single step time cost of CBEngine under the traffic flow of $10^{5}$ vehicles is 0.2610 seconds, while that of SUMO~\cite{SUMO2018} and Cityflow~\cite{zhang2019cityflow} are 8.9111 seconds and 1.1058 seconds, respectively. Specifically, two baselines are not able to run the case with 1,000,000 vehicles. For Cityflow, the reason has been discussed in Section~\ref{sec:lanechanging}.

\subsection{Plausibility}

\begin{figure}[ht]
    \centering
    \includegraphics[width=\linewidth]{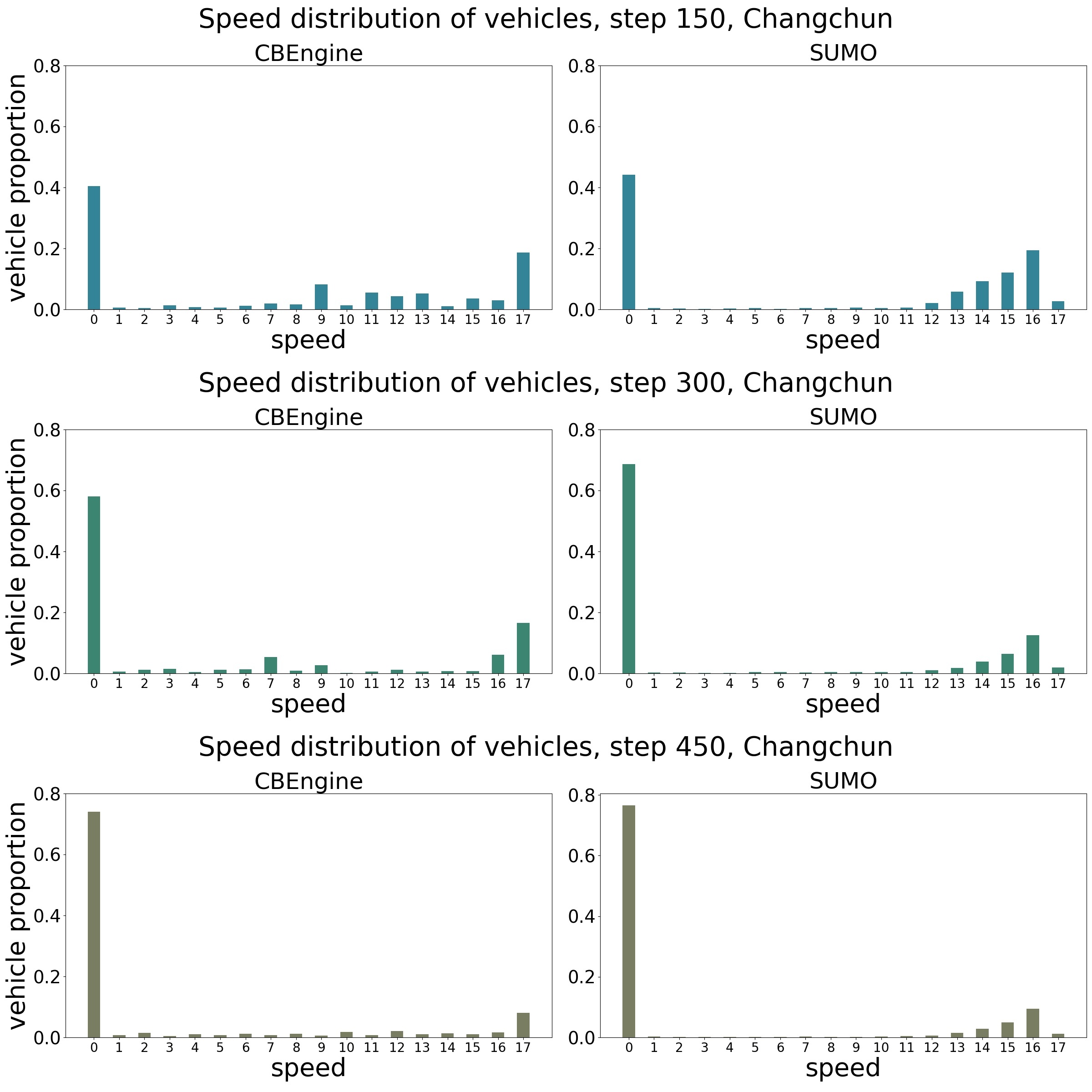}
    \caption{Comparison of speed distributions of vehicles, CBEngine and SUMO}
    \label{fig:sd}
    \vspace{-0.4cm}
\end{figure}

\begin{figure}[th]
    \centering
    \includegraphics[width=\linewidth]{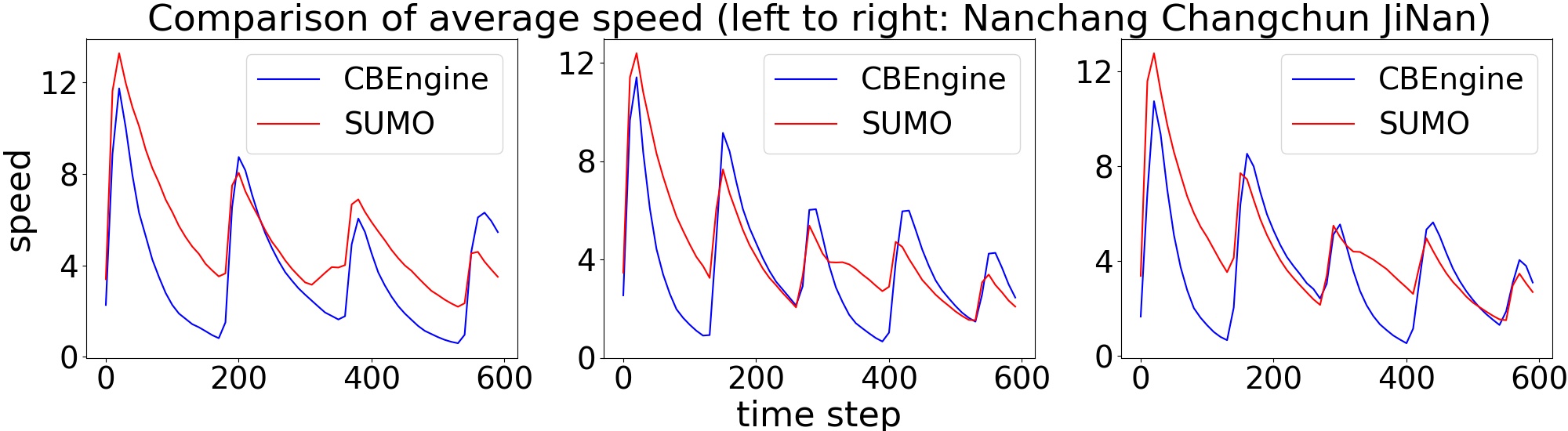}
    \caption{Comparison of the average speed of vehicles, CBEngine and SUMO}
    \label{fig:curve}
    \vspace{-0.3cm}
\end{figure}

Although CBEngine provides driving model customization, which decreases the impact of the default driving model, we conduct two experiments to evaluate its plausibility by comparing the vehicle behavior of CBEngine and that of SUMO~\cite{SUMO2018}. First, we compare the average speed of vehicles in CBEngine and SUMO under the same simulation setups over 600 seconds. The experiments are finished on three experimental setups used in our efficiency and scalability experiments: Nanchang, Changchun, and JiNan. The result is visualized in Figure~\ref{fig:curve}. Overall tendencies of two average speeds fit each other approximately, while that in CBEngine is more volatile. This can be explained by the difference in the driving model and will not have visible impacts on traffic flow statistics, which is the main factor in learning traffic control policies. 

The second experiment focuses on the speed distributions of vehicles in CBEngine and SUMO. We compare the speed distributions in 150 seconds, 300 seconds, and 450 seconds under the experimental setup of Changchun used in our efficiency experiments. The comparison is visualized in Figure~\ref{fig:sd}. Normalized Wasserstein distance between two distributions is shown in Table~\ref{tab:sd}. The two distributions are similar to each other roughly but differ in details, which may not influence the overall performance of the simulated traffic in the level of traffic statistics. Also, the driving model customization supported by CBEngine makes it possible for users to use driving models according to their needs, which greatly improves the plausibility of CBEngine, compared to SUMO and Cityflow.

\begin{table}[htbp]
\begin{tabular}{c|c|c|c}
\toprule
Time Step  & 150 & 300 & 450   \\ 
\midrule
Normalized Wasserstein Distance & 0.0141 & 0.0124 & 0.0089 \\
\bottomrule
\end{tabular}
\caption{Normalized Wasserstein distance between speed distribution of CBEngine and SUMO}
\label{tab:sd}
\vspace{-0.5cm}
\end{table}

\section{CBData: Traffic Data Network Connected to CBEngine}
\label{sec:digital}

\begin{figure*}[htbp]
    \centering
    \includegraphics[width=\linewidth]{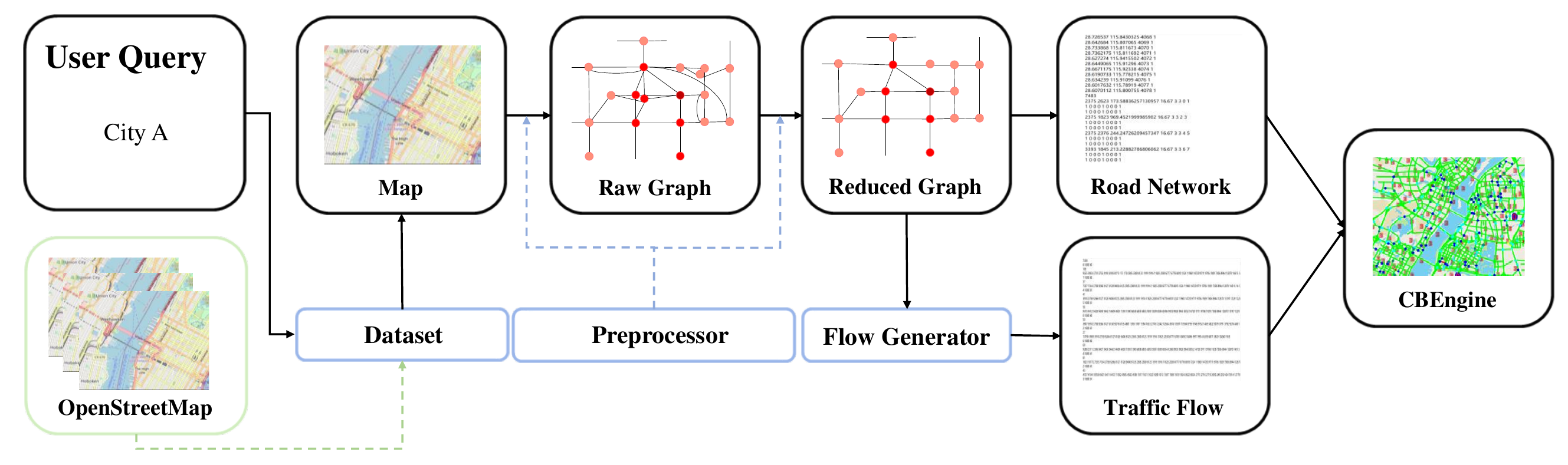}
    \caption{The simulation input data pipeline of CBData.}
    \label{fig:data}
    \vspace{-0.3cm}
\end{figure*}

In this section, we introduce our data tool CBData. CBData serves to provide enriched input data supporting large-scale traffic simulation. The support is achieved by a dataset with raw road networks of 100 main cities all around the world and the transformation pipeline shown in Figure~\ref{fig:data}. Moreover, CBData includes two other pipelines that help the simulator learn from other traffic data.


\begin{figure}[htbp]
    \centering
    \includegraphics[width=\linewidth]{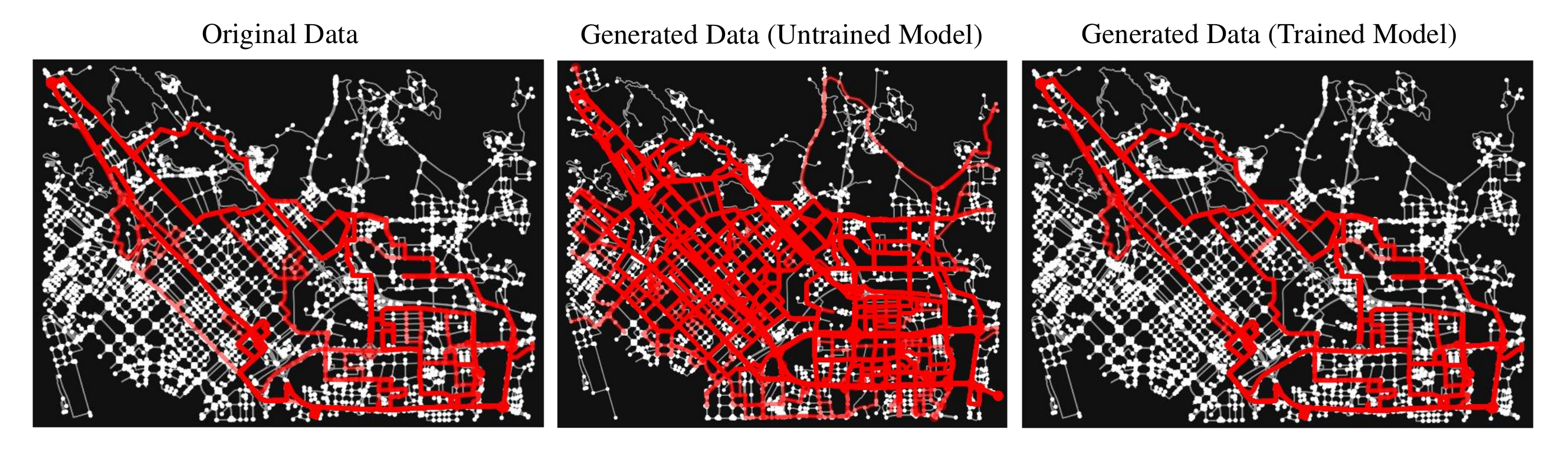}
    \caption{Visualization of the routing result. The deeper red is, the more frequent the route is picked.}
    \label{fig:routing_map}
\end{figure}

\begin{figure}[htbp]
	\centering
	\begin{minipage}{\linewidth}
		\centering
		\includegraphics[width=\linewidth]{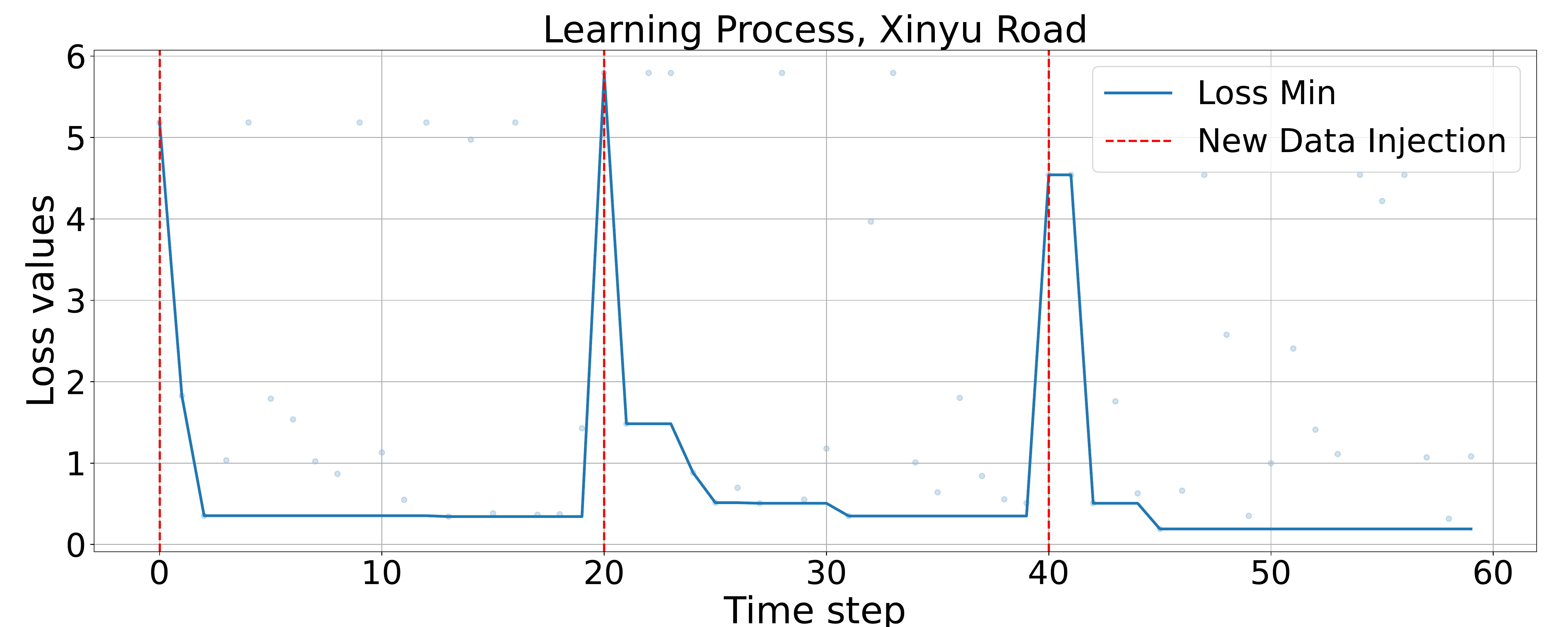}
	\end{minipage}\\
	\begin{minipage}{\linewidth}
		\centering
		\includegraphics[width=\linewidth]{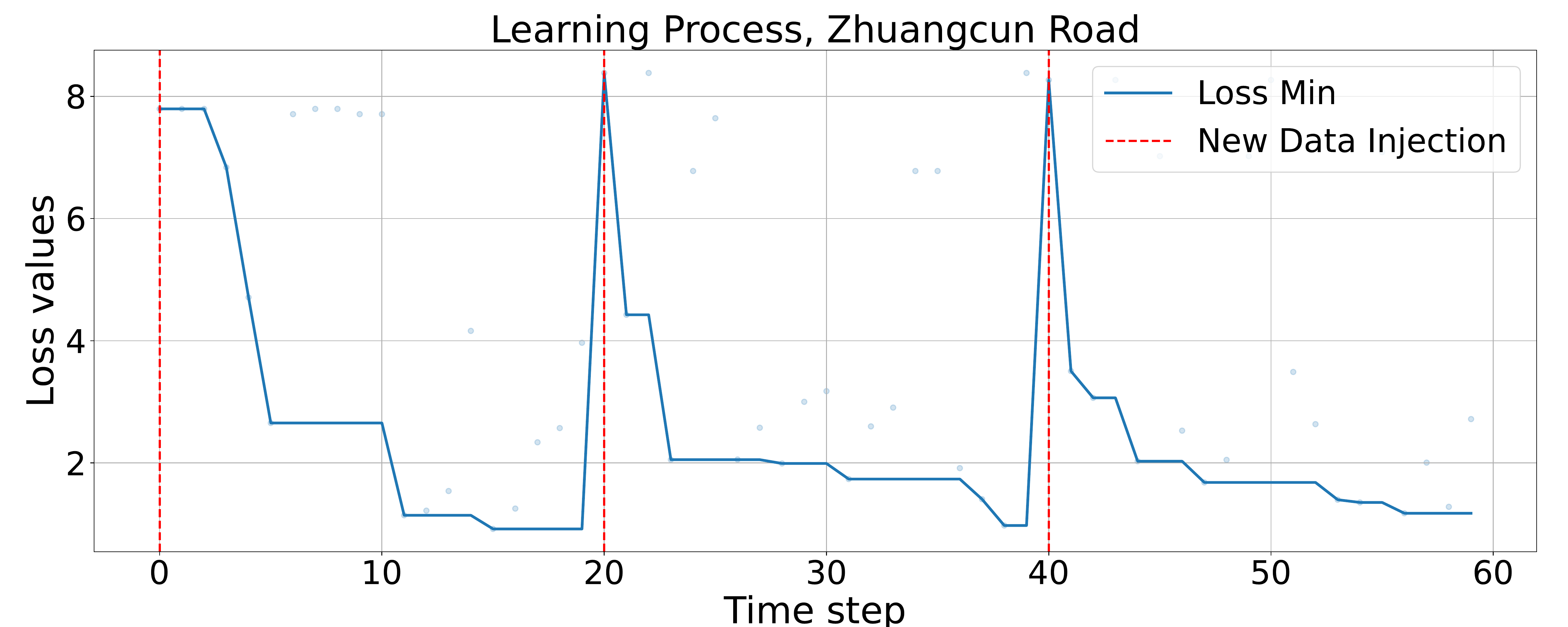}
	\end{minipage}
	\caption{Loss curves during learning the driving module on two roads.}
	\label{fig:data-driven}
 \vspace{-0.2cm}
\end{figure}

\begin{figure}[htbp]
    \centering
    \includegraphics[width=\linewidth]{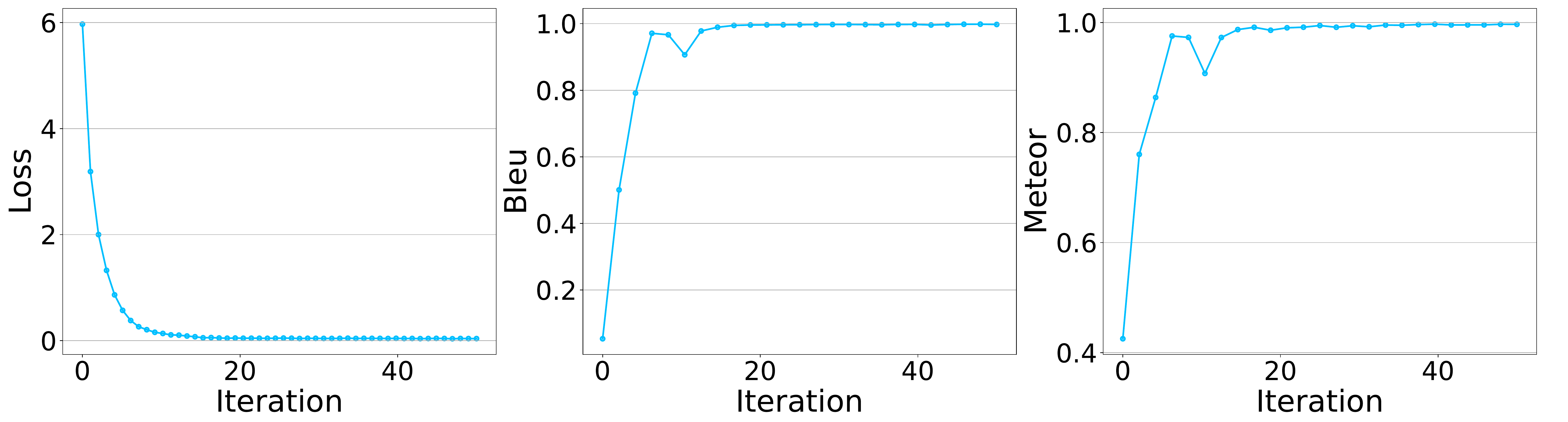}
    \caption{Curves of loss, BLEU, and METEOR during learning the routing module.}
    \label{fig:routing_loss}
    \vspace{-0.3cm}
\end{figure}

\subsection{Pipeline: Simulation Input Data Supporting}

Despite existing traffic simulators being capable of simulating the evolution of urban traffic, the application of traffic simulation is vastly limited by the shortage of input data. Specifically, to start up a simulation, the simulator takes road networks and traffic flows as input data. At present, there is no convenient access to these two kinds of data, although road networks of almost all main cities in the world can be extracted from open-source map data~\cite{haklay2008openstreetmap}.

To disentangle this problem, we implement a pipeline to bridge open-source map data and input data for simulation. This pipeline provides a one-click service to offer enriched input data for large-scale traffic simulation, solving the shortage of input data. 

As shown in Figure~\ref{fig:data}, the pipeline consists of a dataset, a preprocessor, and a flow generator. 

\textbf{Dataset:} We obtain the map data of the whole world from OpenStreetMap~\cite{haklay2008openstreetmap}. We extract the road network data of 100 main cities and store the data in our dataset. The dataset is now available on Google Drive (See Appendix~\ref{appendix:roadnetwork}). Users can directly download the data and pick up their interested road networks. Details of the dataset are given in Appendix~\ref{appendix:dataset}.

\textbf{Preprocessor:} The preprocessor first constructs the road network as a raw graph by matching and connecting edges and nodes. The raw graph is then cleared to remove the redundant nodes and graphs. This is necessary because redundancy is common in open-source map data. After removing the redundancy, the reduced graph is transformed into the road network in the standard format. 

\textbf{Flow Generator:} The flow generator generates the traffic flow for a road network. Given the total number of vehicles, the generator assigns origins and destinations for these vehicles, respectively, which distribute as average as possible. The default route for each pair of the origin and the destination is the shortest path and can be changed by the routing model when running the simulation.


\subsection{Pipeline: Learning to simulate from Traffic Data}

In addition to map data, there is a lot of other traffic data with the potential to enhance the plausibility of traffic simulation. In CBData, we propose two paradigmatic pipelines to illustrate how to learn to simulate from traffic data. Note that we are not to propose effective methods but provide a paradigm for learning to simulate.

\subsubsection{Learning to Simulate Driving}

The driving model determines how drivers accelerate and decelerate according to their observation of the circumstance. The driving behavior in different traffic or different cities can be distinctly different. Therefore, learning the driving parameters of the traffic simulator from the traffic data is sound for traffic simulation. It helps the simulator to behave more plausibly like the local drivers. 

The goal of learning to drive is to find a set of driving model parameters that can minimize the gap between the traffic data and the simulator, $e.g.$ that between the observed speed in the real data and the observed speed in the simulator, with the same traffic flow. As mentioned in Section~\ref{sec:cbenginemodel}, the driving model of CBEngine is easy to modify. Here, we adopt the default model and select three parameters from the model as the parameters to be optimized: acceleration maximum, deceleration maximum, and speed limit. 

We use a black-box optimization toolkit OpenBox~\cite{li2021openbox} as the optimization tool. Note that users can use any other optimization toolkit according to their needs. OpenBox searches for parameters to fill the gap between simulation observation and the ground truth. The 1-hour GPS trajectory data of vehicles on two roads in Shenzhen, China are used to search the parameters. The observation interval is 1 minute. For every 20 minutes, the traffic distribution will change. We expect that acceleration parameters can be continuously learned when the data changes.

The loss curves of the learning process of these two avenues are displayed in Figure~\ref{fig:data-driven}. The gap between the observed speed average and the ground truth is decreasing as time changes. After the traffic changes at the 20th and the 40th minute, the driving model can not fit the new traffic data. Hence, we observe a high loss immediately. After a few minutes, the gap continuously decreases. Note that, the difference still maintains a certain positive value. This implies the potential to change the driving module to pursue a better performance of driving module correction.

\begin{figure*}[hbtp]
    \centering
    \includegraphics[width=0.9\linewidth]{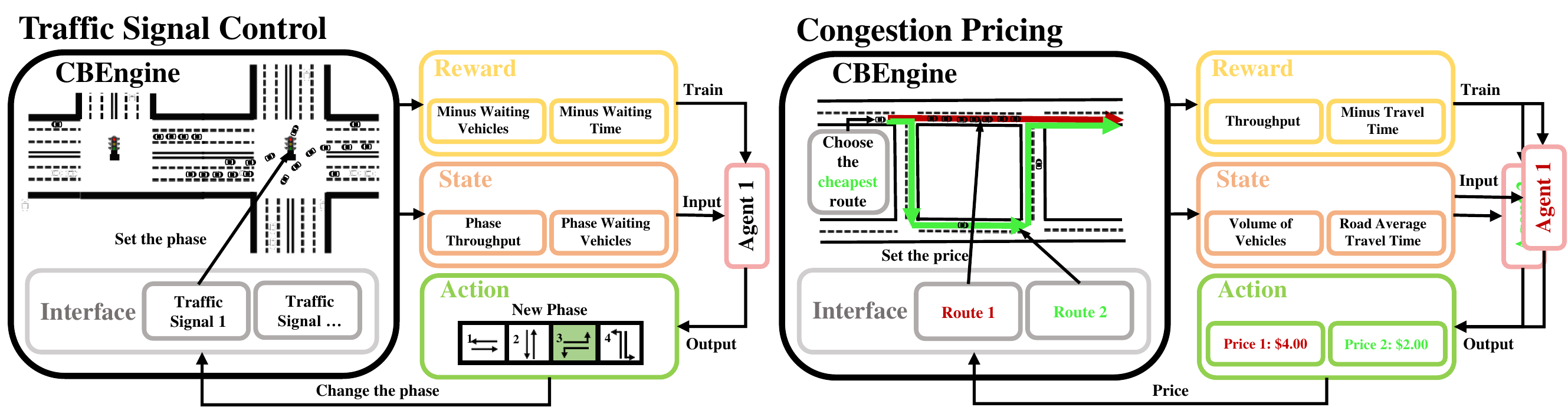}
    \caption{Illustration of two scenarios: traffic signal control and congestion pricing.}
    \label{fig:scenario}
\end{figure*}

\subsubsection{Learning to Simulate Routing}
The routing model determines how vehicles route themselves, given the origin and destination. A data-dependent routing model can be formulated as a route generator, which generates routes for certain origins and destinations based on real trajectories. We exploit the Recurrent Neural Networks (RNNs) as our route generator~\cite{choi2021trajgail}.


We use part of a vehicle trajectory dataset in Shenzhen, China. This dataset contains 22 different routes in total. We train the RNN~\cite{choi2021trajgail} to learn the distribution of routes and conduct routing for vehicles in the simulator. Figure~\ref{fig:routing_loss} shows the result of loss curves and two trajectory generating metrics, BLEU~\cite{bleu} and METEOR~\cite{meteor}. The loss converges to 0 at the first twenty iterations, while BLEU and METEOR get close to 1. This indicates that at this stage, routes generated overlap at least one route in the real trajectory data. 

In addition, the routing result is visualized in Figure~\ref{fig:routing_map}. Compared with an untrained generator, the trained generator recognizes the main distribution pattern of real trajectories. Meanwhile, it tends to ignore some infrequently-appearing routes.

\section{CBScenario: Environment for Large-Scale traffic control policies}
\label{sec:scenario}

\begin{table*}[hbpt]

\setlength{\tabcolsep}{5mm}{
\begin{tabular}{c|c|c|c|c}
\toprule
Dataset   & \multicolumn{2}{|c}{Hangzhou} & \multicolumn{2}{|c}{Manhattan}\\ \midrule
\midrule
Metrics  & Throughput & Travel Time(s) & Throughput & Travel Time(s)\\ 
\midrule
FixedTime  & 2184 & 1478.01 & 2894 & 1309.88   \\  
MaxPressure & 3336 & 700.66 & 3364 & 805.14     \\
SOTL  & 1122 & 305.79 & 137 &  488.62   \\
DQN  & 3573 & 309.10 & 3926 &  375.51  \\
\bottomrule
\end{tabular}}
\setlength{\tabcolsep}{5mm}{
\begin{tabular}{c|c|c|c|c}
\toprule
Dataset   & \multicolumn{2}{c|}{Hangzhou} & \multicolumn{2}{c}{Manhattan}\\ \midrule
\midrule
Metrics  & Throughput & Travel Time(s) & Throughput & Travel Time(s)   \\ 
\midrule
No-Change & 2176 & 1455 & 2911 & 1328 \\
Random & 3008 & 644.00 & 3459 & 1120.69 \\  
Deltatoll & 3186 & 604.00 & 3670 & 960.17  \\
EBGtoll & 2803  & 310.18 & 3494 & 1019.85 \\
\bottomrule
\end{tabular}
}
\caption{Performance of baseline algorithms on traffic signal control (Up) and congestion pricing (Down).}
\label{tab:tsc}
\label{tab:cp}
\vspace{-0.5cm}
\end{table*}


In this section, we introduce CBScenario, an interactive environment for large-scale traffic control policies. CBScenario benefits from the large-scale traffic simulation supported by CBEngine and CBData and is capable of training traffic control policies for city-level traffic. Concretely, CBScenario includes benchmarks for two traffic control policies: "Traffic Signal Control" and "Congestion Pricing". We conduct experiments on our environment to show the plausibility of the traffic simulation.

\subsection{Traffic Control Policy 1: Traffic Signal Control}

The traffic signal control problem~\cite{wei2019survey} tries to improve the performance of urban traffic by carefully choosing the phase of traffic signals at intersections. An ideal traffic signal control policy can capture the global and local traffic dynamics and allocate more passing time to the phase with higher traffic pressure. Figure~\ref{fig:scenario} (left) shows the problem setting of traffic signal control.

In consideration of the Markov nature of traffic signal control, the traffic signal control can be formulated as a Markov Decision Process (MDP):
\begin{itemize}
    \item \textbf{State:} Intersection-level and road-level observation and statistics of observation, $e.g.$ the number of waiting vehicles on the road, historical average vehicle throughput of different phases at the intersection. 
    \item \textbf{Action:} Decide which directions can pass.
    \item \textbf{Reward:} Metrics measuring the performance of urban traffic. We provide two widely-used metrics: the total number of waiting vehicles at the intersections and the average waiting time during an action interval.
\end{itemize}
We implement several baseline algorithms to justify the plausibility of our simulation. Two metrics are used to evaluate their performances: arriving vehicle throughput and average travel time of vehicles. The results are shown in Table~\ref{tab:tsc}. Two transportation methods, MaxPressure~\cite{varaiya2013max} and Self Organized Traffic Light (SOTL)~\cite{cools2013self}, show a degree of advantages in 
increasing throughput and reducing travel time. As a learning-based traffic control policy, DQN~\cite{mnih2015human,wei2018intellilight} performs even better. See Appendix~\ref{appendix:exp} for more details.

\subsection{Traffic Control Policy 2: Congestion Pricing}

Congestion pricing reroutes vehicles by dynamically assigning prices to different routes and guiding vehicles to drive on the route of the lowest price. Good pricing methods tend to allocate heavy traffic on different routes with a trade-off of the distance and the capacity, therefore enhancing traffic efficiency. Figure~\ref{fig:scenario} (right) shows the setting of congestion pricing.

Similar to traffic signal control, we define congestion pricing as an MDP:
\begin{itemize}
\vspace{-0.1cm}
\item \textbf{State:} Road-level observation and statistics of observation. For the observation commonly used, we have the vehicle number and the average speed of vehicles on the road. 
    \item \textbf{Action:} Price roads in the road network.
    \item \textbf{Reward:} Metrics measuring the performance of urban traffic. A common reward is the average travel distance of vehicles during an interval. 
\end{itemize}
Congestion pricing has been researched in the transportation field for a while yet few studies use data-driven methods. We implement two transportation-based methods, Random and Deltatoll~\cite{dtoll}, and an RL algorithm, EBGtoll~\cite{EBGtoll}. More details of the experiment are given in Appendix~\ref{appendix:exp}. Results are shown in Table~\ref{tab:cp}. Random outperforms No-change which keeps original routes for vehicles. This is plausible because random rerouting averagely allocates traffic on all available routes. Deltatoll and EBGtoll behave similarly and outperform Random in both evaluation metrics.

\section{Conclusion} 
In this paper, we present CBLab, a toolkit for scalable traffic simulation. CBLab provides the first simulator to support real-time simulation on large-scale cities with more than 10,000 intersections and 1,000,000 vehicles. A data tool is implemented to supply large-scale simulation data. We build an interactive environment and benchmarks for two traffic control policies. CBLab is the first infrastructure supporting the training of large-scale traffic control policies. 

\section*{Acknowledgement}
\label{sec:ack}
This work was sponsored by National Key Research and Development Program of China under Grant No.2022YFB3904204, National Natural Science Foundation of China under Grant No. 62102246, 62272301, and Provincial Key Research and Development Program of Zhejiang under Grant No. 2021C01034.
\bibliographystyle{ACM-Reference-Format}
\bibliography{sample-base}


\begin{thebibliography}{30}


\ifx \showCODEN    \undefined \def \showCODEN     #1{\unskip}     \fi
\ifx \showDOI      \undefined \def \showDOI       #1{#1}\fi
\ifx \showISBNx    \undefined \def \showISBNx     #1{\unskip}     \fi
\ifx \showISBNxiii \undefined \def \showISBNxiii  #1{\unskip}     \fi
\ifx \showISSN     \undefined \def \showISSN      #1{\unskip}     \fi
\ifx \showLCCN     \undefined \def \showLCCN      #1{\unskip}     \fi
\ifx \shownote     \undefined \def \shownote      #1{#1}          \fi
\ifx \showarticletitle \undefined \def \showarticletitle #1{#1}   \fi
\ifx \showURL      \undefined \def \showURL       {\relax}        \fi
\providecommand\bibfield[2]{#2}
\providecommand\bibinfo[2]{#2}
\providecommand\natexlab[1]{#1}
\providecommand\showeprint[2][]{arXiv:#2}

\bibitem[Banerjee and Lavie(2005)]%
        {meteor}
\bibfield{author}{\bibinfo{person}{Satanjeev Banerjee} {and}
  \bibinfo{person}{Alon Lavie}.} \bibinfo{year}{2005}\natexlab{}.
\newblock \showarticletitle{METEOR: An Automatic Metric for MT Evaluation with
  Improved Correlation with Human Judgments}. In
  \bibinfo{booktitle}{\emph{IEEvaluation@ACL}}. \bibinfo{pages}{65--72}.
\newblock
\urldef\tempurl%
\url{https://aclanthology.info/papers/W05-0909/w05-0909}
\showURL{%
\tempurl}


\bibitem[Chen et~al\mbox{.}(2020a)]%
        {chen2020toward}
\bibfield{author}{\bibinfo{person}{Chacha Chen}, \bibinfo{person}{Hua Wei},
  \bibinfo{person}{Nan Xu}, \bibinfo{person}{Guanjie Zheng},
  \bibinfo{person}{Ming Yang}, \bibinfo{person}{Yuanhao Xiong},
  \bibinfo{person}{Kai Xu}, {and} \bibinfo{person}{Zhenhui Li}.}
  \bibinfo{year}{2020}\natexlab{a}.
\newblock \showarticletitle{Toward a thousand lights: Decentralized deep
  reinforcement learning for large-scale traffic signal control}. In
  \bibinfo{booktitle}{\emph{Proceedings of the AAAI Conference on Artificial
  Intelligence}}, Vol.~\bibinfo{volume}{34}. \bibinfo{pages}{3414--3421}.
\newblock


\bibitem[Chen et~al\mbox{.}(2020b)]%
        {qarsumo}
\bibfield{author}{\bibinfo{person}{Hao Chen}, \bibinfo{person}{Ke Yang},
  \bibinfo{person}{Stefano~Giovanni Rizzo}, \bibinfo{person}{Giovanna Vantini},
  \bibinfo{person}{Phillip Taylor}, \bibinfo{person}{Xiaosong Ma}, {and}
  \bibinfo{person}{Sanjay Chawla}.} \bibinfo{year}{2020}\natexlab{b}.
\newblock \showarticletitle{QarSUMO: A Parallel, Congestion-Optimized Traffic
  Simulator}. In \bibinfo{booktitle}{\emph{Proceedings of the 28th
  International Conference on Advances in Geographic Information Systems}}.
  \bibinfo{pages}{578–588}.
\newblock


\bibitem[Choi et~al\mbox{.}(2021)]%
        {choi2021trajgail}
\bibfield{author}{\bibinfo{person}{Seongjin Choi}, \bibinfo{person}{Jiwon Kim},
  {and} \bibinfo{person}{Hwasoo Yeo}.} \bibinfo{year}{2021}\natexlab{}.
\newblock \showarticletitle{TrajGAIL: Generating urban vehicle trajectories
  using generative adversarial imitation learning}.
\newblock \bibinfo{journal}{\emph{Transportation Research Part C: Emerging
  Technologies}}  \bibinfo{volume}{128} (\bibinfo{year}{2021}),
  \bibinfo{pages}{103091}.
\newblock


\bibitem[Cools et~al\mbox{.}(2013)]%
        {cools2013self}
\bibfield{author}{\bibinfo{person}{Seung-Bae Cools}, \bibinfo{person}{Carlos
  Gershenson}, {and} \bibinfo{person}{Bart D’Hooghe}.}
  \bibinfo{year}{2013}\natexlab{}.
\newblock \showarticletitle{Self-organizing traffic lights: A realistic
  simulation}.
\newblock In \bibinfo{booktitle}{\emph{Advances in applied self-organizing
  systems}}. \bibinfo{publisher}{Springer}, \bibinfo{pages}{45--55}.
\newblock


\bibitem[Haarnoja et~al\mbox{.}(2018)]%
        {haarnoja2018soft}
\bibfield{author}{\bibinfo{person}{Tuomas Haarnoja}, \bibinfo{person}{Aurick
  Zhou}, \bibinfo{person}{Pieter Abbeel}, {and} \bibinfo{person}{Sergey
  Levine}.} \bibinfo{year}{2018}\natexlab{}.
\newblock \showarticletitle{Soft actor-critic: Off-policy maximum entropy deep
  reinforcement learning with a stochastic actor}. In
  \bibinfo{booktitle}{\emph{International conference on machine learning}}.
  PMLR, \bibinfo{pages}{1861--1870}.
\newblock


\bibitem[Haklay and Weber(2008)]%
        {haklay2008openstreetmap}
\bibfield{author}{\bibinfo{person}{Mordechai Haklay} {and}
  \bibinfo{person}{Patrick Weber}.} \bibinfo{year}{2008}\natexlab{}.
\newblock \showarticletitle{Openstreetmap: User-generated street maps}.
\newblock \bibinfo{journal}{\emph{IEEE Pervasive computing}}
  \bibinfo{volume}{7}, \bibinfo{number}{4} (\bibinfo{year}{2008}),
  \bibinfo{pages}{12--18}.
\newblock


\bibitem[Krau{\ss}(1998)]%
        {krauss1998microscopic}
\bibfield{author}{\bibinfo{person}{Stefan Krau{\ss}}.}
  \bibinfo{year}{1998}\natexlab{}.
\newblock \showarticletitle{Microscopic modeling of traffic flow: Investigation
  of collision free vehicle dynamics}.
\newblock  (\bibinfo{year}{1998}).
\newblock


\bibitem[Li et~al\mbox{.}(2021)]%
        {li2021openbox}
\bibfield{author}{\bibinfo{person}{Yang Li}, \bibinfo{person}{Yu Shen},
  \bibinfo{person}{Wentao Zhang}, \bibinfo{person}{Yuanwei Chen},
  \bibinfo{person}{Huaijun Jiang}, \bibinfo{person}{Mingchao Liu},
  \bibinfo{person}{Jiawei Jiang}, \bibinfo{person}{Jinyang Gao},
  \bibinfo{person}{Wentao Wu}, \bibinfo{person}{Zhi Yang}, {et~al\mbox{.}}}
  \bibinfo{year}{2021}\natexlab{}.
\newblock \showarticletitle{Openbox: A generalized black-box optimization
  service}. In \bibinfo{booktitle}{\emph{Proceedings of the 27th ACM SIGKDD
  Conference on Knowledge Discovery \& Data Mining}}.
  \bibinfo{pages}{3209--3219}.
\newblock


\bibitem[Lopez et~al\mbox{.}(2018)]%
        {SUMO2018}
\bibfield{author}{\bibinfo{person}{Pablo~Alvarez Lopez},
  \bibinfo{person}{Michael Behrisch}, \bibinfo{person}{Laura Bieker-Walz},
  \bibinfo{person}{Jakob Erdmann}, \bibinfo{person}{Yun-Pang
  Fl{\"o}tter{\"o}d}, \bibinfo{person}{Robert Hilbrich},
  \bibinfo{person}{Leonhard L{\"u}cken}, \bibinfo{person}{Johannes Rummel},
  \bibinfo{person}{Peter Wagner}, {and} \bibinfo{person}{Evamarie
  Wie{\ss}ner}.} \bibinfo{year}{2018}\natexlab{}.
\newblock \showarticletitle{Microscopic Traffic Simulation using SUMO}, In
  \bibinfo{booktitle}{The 21st IEEE International Conference on Intelligent
  Transportation Systems}.
\newblock \bibinfo{journal}{\emph{IEEE Intelligent Transportation Systems
  Conference (ITSC)}}.
\newblock
\urldef\tempurl%
\url{https://elib.dlr.de/124092/}
\showURL{%
\tempurl}


\bibitem[Mirzaei et~al\mbox{.}(2018)]%
        {edtoll}
\bibfield{author}{\bibinfo{person}{Hamid Mirzaei}, \bibinfo{person}{Guni
  Sharon}, \bibinfo{person}{Stephen Boyles}, \bibinfo{person}{Tony Givargis},
  {and} \bibinfo{person}{Peter Stone}.} \bibinfo{year}{2018}\natexlab{}.
\newblock \showarticletitle{Enhanced delta-tolling: Traffic optimization via
  policy gradient reinforcement learning}. In \bibinfo{booktitle}{\emph{2018
  21st International Conference on Intelligent Transportation Systems (ITSC)}}.
  IEEE, \bibinfo{pages}{47--52}.
\newblock


\bibitem[Mnih et~al\mbox{.}(2015)]%
        {mnih2015human}
\bibfield{author}{\bibinfo{person}{Volodymyr Mnih}, \bibinfo{person}{Koray
  Kavukcuoglu}, \bibinfo{person}{David Silver}, \bibinfo{person}{Andrei~A
  Rusu}, \bibinfo{person}{Joel Veness}, \bibinfo{person}{Marc~G Bellemare},
  \bibinfo{person}{Alex Graves}, \bibinfo{person}{Martin Riedmiller},
  \bibinfo{person}{Andreas~K Fidjeland}, \bibinfo{person}{Georg Ostrovski},
  {et~al\mbox{.}}} \bibinfo{year}{2015}\natexlab{}.
\newblock \showarticletitle{Human-level control through deep reinforcement
  learning}.
\newblock \bibinfo{journal}{\emph{nature}} \bibinfo{volume}{518},
  \bibinfo{number}{7540} (\bibinfo{year}{2015}), \bibinfo{pages}{529--533}.
\newblock


\bibitem[Oroojlooy et~al\mbox{.}(2020)]%
        {oroojlooy2020attendlight}
\bibfield{author}{\bibinfo{person}{Afshin Oroojlooy},
  \bibinfo{person}{Mohammadreza Nazari}, \bibinfo{person}{Davood Hajinezhad},
  {and} \bibinfo{person}{Jorge Silva}.} \bibinfo{year}{2020}\natexlab{}.
\newblock \showarticletitle{Attendlight: Universal attention-based
  reinforcement learning model for traffic signal control}.
\newblock \bibinfo{journal}{\emph{Advances in Neural Information Processing
  Systems}}  \bibinfo{volume}{33} (\bibinfo{year}{2020}),
  \bibinfo{pages}{4079--4090}.
\newblock


\bibitem[Papineni et~al\mbox{.}(2002)]%
        {bleu}
\bibfield{author}{\bibinfo{person}{Kishore Papineni}, \bibinfo{person}{Salim
  Roukos}, \bibinfo{person}{Todd Ward}, {and} \bibinfo{person}{Wei-Jing Zhu}.}
  \bibinfo{year}{2002}\natexlab{}.
\newblock \showarticletitle{Bleu: a Method for Automatic Evaluation of Machine
  Translation}. In \bibinfo{booktitle}{\emph{ACL}}. \bibinfo{pages}{311--318}.
\newblock
\urldef\tempurl%
\url{http://www.aclweb.org/anthology/P02-1040.pdf}
\showURL{%
\tempurl}


\bibitem[Qiu et~al\mbox{.}(2019)]%
        {EBGtoll}
\bibfield{author}{\bibinfo{person}{Wei Qiu}, \bibinfo{person}{Haipeng Chen},
  {and} \bibinfo{person}{Bo An}.} \bibinfo{year}{2019}\natexlab{}.
\newblock \showarticletitle{Dynamic Electronic Toll Collection via Multi-Agent
  Deep Reinforcement Learning with Edge-Based Graph Convolutional Networks.}.
  In \bibinfo{booktitle}{\emph{IJCAI}}. \bibinfo{pages}{4568--4574}.
\newblock


\bibitem[Sharon et~al\mbox{.}(2017)]%
        {dtoll}
\bibfield{author}{\bibinfo{person}{Guni Sharon}, \bibinfo{person}{Michael~W
  Levin}, \bibinfo{person}{Josiah~P Hanna}, \bibinfo{person}{Tarun Rambha},
  \bibinfo{person}{Stephen~D Boyles}, {and} \bibinfo{person}{Peter Stone}.}
  \bibinfo{year}{2017}\natexlab{}.
\newblock \showarticletitle{Network-wide adaptive tolling for connected and
  automated vehicles}.
\newblock \bibinfo{journal}{\emph{Transportation Research Part C: Emerging
  Technologies}}  \bibinfo{volume}{84} (\bibinfo{year}{2017}),
  \bibinfo{pages}{142--157}.
\newblock


\bibitem[Varaiya(2013)]%
        {varaiya2013max}
\bibfield{author}{\bibinfo{person}{Pravin Varaiya}.}
  \bibinfo{year}{2013}\natexlab{}.
\newblock \showarticletitle{Max pressure control of a network of signalized
  intersections}.
\newblock \bibinfo{journal}{\emph{Transportation Research Part C: Emerging
  Technologies}}  \bibinfo{volume}{36} (\bibinfo{year}{2013}),
  \bibinfo{pages}{177--195}.
\newblock


\bibitem[Wei et~al\mbox{.}(2019a)]%
        {wei2019presslight}
\bibfield{author}{\bibinfo{person}{Hua Wei}, \bibinfo{person}{Chacha Chen},
  \bibinfo{person}{Guanjie Zheng}, \bibinfo{person}{Kan Wu},
  \bibinfo{person}{Vikash Gayah}, \bibinfo{person}{Kai Xu}, {and}
  \bibinfo{person}{Zhenhui Li}.} \bibinfo{year}{2019}\natexlab{a}.
\newblock \showarticletitle{Presslight: Learning max pressure control to
  coordinate traffic signals in arterial network}. In
  \bibinfo{booktitle}{\emph{Proceedings of the 25th ACM SIGKDD International
  Conference on Knowledge Discovery \& Data Mining}}.
  \bibinfo{pages}{1290--1298}.
\newblock


\bibitem[Wei et~al\mbox{.}(2019b)]%
        {wei2019colight}
\bibfield{author}{\bibinfo{person}{Hua Wei}, \bibinfo{person}{Nan Xu},
  \bibinfo{person}{Huichu Zhang}, \bibinfo{person}{Guanjie Zheng},
  \bibinfo{person}{Xinshi Zang}, \bibinfo{person}{Chacha Chen},
  \bibinfo{person}{Weinan Zhang}, \bibinfo{person}{Yanmin Zhu},
  \bibinfo{person}{Kai Xu}, {and} \bibinfo{person}{Zhenhui Li}.}
  \bibinfo{year}{2019}\natexlab{b}.
\newblock \showarticletitle{Colight: Learning network-level cooperation for
  traffic signal control}. In \bibinfo{booktitle}{\emph{Proceedings of the 28th
  ACM International Conference on Information and Knowledge Management}}.
  \bibinfo{pages}{1913--1922}.
\newblock


\bibitem[Wei et~al\mbox{.}(2019c)]%
        {wei2019survey}
\bibfield{author}{\bibinfo{person}{Hua Wei}, \bibinfo{person}{Guanjie Zheng},
  \bibinfo{person}{Vikash Gayah}, {and} \bibinfo{person}{Zhenhui Li}.}
  \bibinfo{year}{2019}\natexlab{c}.
\newblock \showarticletitle{A survey on traffic signal control methods}.
\newblock \bibinfo{journal}{\emph{arXiv preprint arXiv:1904.08117}}
  (\bibinfo{year}{2019}).
\newblock


\bibitem[Wei et~al\mbox{.}(2018)]%
        {wei2018intellilight}
\bibfield{author}{\bibinfo{person}{Hua Wei}, \bibinfo{person}{Guanjie Zheng},
  \bibinfo{person}{Huaxiu Yao}, {and} \bibinfo{person}{Zhenhui Li}.}
  \bibinfo{year}{2018}\natexlab{}.
\newblock \showarticletitle{Intellilight: A reinforcement learning approach for
  intelligent traffic light control}. In \bibinfo{booktitle}{\emph{Proceedings
  of the 24th ACM SIGKDD International Conference on Knowledge Discovery \&
  Data Mining}}. \bibinfo{pages}{2496--2505}.
\newblock


\bibitem[Wu et~al\mbox{.}(2021)]%
        {wu2021dynstgat}
\bibfield{author}{\bibinfo{person}{Libing Wu}, \bibinfo{person}{Min Wang},
  \bibinfo{person}{Dan Wu}, {and} \bibinfo{person}{Jia Wu}.}
  \bibinfo{year}{2021}\natexlab{}.
\newblock \showarticletitle{DynSTGAT: Dynamic Spatial-Temporal Graph Attention
  Network for Traffic Signal Control}. In \bibinfo{booktitle}{\emph{Proceedings
  of the 30th ACM International Conference on Information \& Knowledge
  Management}}. \bibinfo{pages}{2150--2159}.
\newblock


\bibitem[Xu et~al\mbox{.}(2021)]%
        {xu2021hierarchically}
\bibfield{author}{\bibinfo{person}{Bingyu Xu}, \bibinfo{person}{Yaowei Wang},
  \bibinfo{person}{Zhaozhi Wang}, \bibinfo{person}{Huizhu Jia}, {and}
  \bibinfo{person}{Zongqing Lu}.} \bibinfo{year}{2021}\natexlab{}.
\newblock \showarticletitle{Hierarchically and cooperatively learning traffic
  signal control}. In \bibinfo{booktitle}{\emph{Proceedings of the AAAI
  Conference on Artificial Intelligence}}, Vol.~\bibinfo{volume}{35}.
  \bibinfo{pages}{669--677}.
\newblock


\bibitem[Yuan et~al\mbox{.}(2011)]%
        {yuan2011driving}
\bibfield{author}{\bibinfo{person}{Jing Yuan}, \bibinfo{person}{Yu Zheng},
  \bibinfo{person}{Xing Xie}, {and} \bibinfo{person}{Guangzhong Sun}.}
  \bibinfo{year}{2011}\natexlab{}.
\newblock \showarticletitle{Driving with knowledge from the physical world}. In
  \bibinfo{booktitle}{\emph{Proceedings of the 17th ACM SIGKDD international
  conference on Knowledge discovery and data mining}}.
  \bibinfo{pages}{316--324}.
\newblock


\bibitem[Yuan et~al\mbox{.}(2010)]%
        {yuan2010t}
\bibfield{author}{\bibinfo{person}{Jing Yuan}, \bibinfo{person}{Yu Zheng},
  \bibinfo{person}{Chengyang Zhang}, \bibinfo{person}{Wenlei Xie},
  \bibinfo{person}{Xing Xie}, \bibinfo{person}{Guangzhong Sun}, {and}
  \bibinfo{person}{Yan Huang}.} \bibinfo{year}{2010}\natexlab{}.
\newblock \showarticletitle{T-drive: driving directions based on taxi
  trajectories}. In \bibinfo{booktitle}{\emph{Proceedings of the 18th
  SIGSPATIAL International conference on advances in geographic information
  systems}}. \bibinfo{pages}{99--108}.
\newblock


\bibitem[Zang et~al\mbox{.}(2020)]%
        {zang2020metalight}
\bibfield{author}{\bibinfo{person}{Xinshi Zang}, \bibinfo{person}{Huaxiu Yao},
  \bibinfo{person}{Guanjie Zheng}, \bibinfo{person}{Nan Xu},
  \bibinfo{person}{Kai Xu}, {and} \bibinfo{person}{Zhenhui Li}.}
  \bibinfo{year}{2020}\natexlab{}.
\newblock \showarticletitle{Metalight: Value-based meta-reinforcement learning
  for traffic signal control}. In \bibinfo{booktitle}{\emph{Proceedings of the
  AAAI Conference on Artificial Intelligence}}, Vol.~\bibinfo{volume}{34}.
  \bibinfo{pages}{1153--1160}.
\newblock


\bibitem[Zhang et~al\mbox{.}(2019)]%
        {zhang2019cityflow}
\bibfield{author}{\bibinfo{person}{Huichu Zhang}, \bibinfo{person}{Siyuan
  Feng}, \bibinfo{person}{Chang Liu}, \bibinfo{person}{Yaoyao Ding},
  \bibinfo{person}{Yichen Zhu}, \bibinfo{person}{Zihan Zhou},
  \bibinfo{person}{Weinan Zhang}, \bibinfo{person}{Yong Yu},
  \bibinfo{person}{Haiming Jin}, {and} \bibinfo{person}{Zhenhui Li}.}
  \bibinfo{year}{2019}\natexlab{}.
\newblock \showarticletitle{Cityflow: A multi-agent reinforcement learning
  environment for large scale city traffic scenario}. In
  \bibinfo{booktitle}{\emph{The world wide web conference}}.
  \bibinfo{pages}{3620--3624}.
\newblock


\bibitem[Zhang et~al\mbox{.}(2020)]%
        {zhang2020generalight}
\bibfield{author}{\bibinfo{person}{Huichu Zhang}, \bibinfo{person}{Chang Liu},
  \bibinfo{person}{Weinan Zhang}, \bibinfo{person}{Guanjie Zheng}, {and}
  \bibinfo{person}{Yong Yu}.} \bibinfo{year}{2020}\natexlab{}.
\newblock \showarticletitle{Generalight: Improving environment generalization
  of traffic signal control via meta reinforcement learning}. In
  \bibinfo{booktitle}{\emph{Proceedings of the 29th ACM International
  Conference on Information \& Knowledge Management}}.
  \bibinfo{pages}{1783--1792}.
\newblock


\bibitem[Zhang et~al\mbox{.}(2022)]%
        {zhang2022expression}
\bibfield{author}{\bibinfo{person}{Liang Zhang}, \bibinfo{person}{Qiang Wu},
  \bibinfo{person}{Jun Shen}, \bibinfo{person}{Linyuan L{\"u}},
  \bibinfo{person}{Bo Du}, {and} \bibinfo{person}{Jianqing Wu}.}
  \bibinfo{year}{2022}\natexlab{}.
\newblock \showarticletitle{Expression might be enough: representing pressure
  and demand for reinforcement learning based traffic signal control}. In
  \bibinfo{booktitle}{\emph{International Conference on Machine Learning}}.
  PMLR, \bibinfo{pages}{26645--26654}.
\newblock


\bibitem[Zheng et~al\mbox{.}(2019)]%
        {zheng2019learning}
\bibfield{author}{\bibinfo{person}{Guanjie Zheng}, \bibinfo{person}{Yuanhao
  Xiong}, \bibinfo{person}{Xinshi Zang}, \bibinfo{person}{Jie Feng},
  \bibinfo{person}{Hua Wei}, \bibinfo{person}{Huichu Zhang},
  \bibinfo{person}{Yong Li}, \bibinfo{person}{Kai Xu}, {and}
  \bibinfo{person}{Zhenhui Li}.} \bibinfo{year}{2019}\natexlab{}.
\newblock \showarticletitle{Learning phase competition for traffic signal
  control}. In \bibinfo{booktitle}{\emph{Proceedings of the 28th ACM
  International Conference on Information and Knowledge Management}}.
  \bibinfo{pages}{1963--1972}.
\newblock


\end{thebibliography}

\appendix

\appendix
\newpage
\section{Key Information of CBLab}
\label{appendix:info}
\subsection{Licensing} 

CBLab uses the MIT license.

\subsection{Code}
\label{appendix:code}

The code of CBLab is available on GitHub.  \\
\url{https://github.com/CityBrainLab/CityBrainLab.git}

\subsection{Documentation}
\label{appendix:doc}

The documentation of CBLab is available. \\
\url{https://cblab-documentation.readthedocs.io/en/latest/}

\subsection{Road Network Dataset}
\label{appendix:roadnetwork}
The road network dataset is available on Google Drive:\\
\url{https://drive.google.com/drive/folders/1q0StZW9ERwMOKQRIT_29ZR-hhGT013J5?usp=sharing}

\subsection{Limitations \& Future works}

CBLab aims to provide efficient and data-driven traffic simulation for research in traffic policies. Although we provide interfaces for using data to optimize the simulator, we cannot obtain all real traffic data from different cities to conduct complete optimization on our simulator. We will invite future contributors to make more traffic data compatible with these interfaces so that they can be used to optimize the simulator and run the simulation. 

Moreover, we plan to implement more scenarios for traffic policies, $e.g.$ traffic restriction. Also, we aim to include more powerful algorithms in these scenarios, $e.g.$ Intellilight and FRAP for traffic signal control. It is our hope that CBLab can serve as an online benchmark for various scenarios with state-of-the-art algorithms in addition to supporting powerful traffic simulation. 

\subsection{Potential Negative Social Impacts}

All traffic simulators will suffer from the gap between simulated and simulated observations. This gap may lead to biased observations and traffic policies based on the simulation. CBLab releases the first traffic simulator open for users to optimize so that it tends to perform as the real traffic. However, the traffic patterns from different places are different. It is worth considering how to ensure the reality of our traffic simulation when simulating the traffic from different places. Hence, we are collecting more traffic data and studying the common patterns among the data, with the hope of trying to resolve this problem.

\section{Details of Experimental Setup}
\label{appendix:exp}
\subsection{Efficiency and Scalability (Section 2.5)}
\begin{table*}[htbp]

\begin{tabular}{c|c|c|c|c|c|c}
\toprule
Dataset   & Nanchang & Changchun & JiNan & Shenzhen & Hangzhou & Shanghai\\ \midrule
\midrule
 & \multicolumn{6}{c}{Time Cost} \\ \midrule
SUMO  & 1239.93 & 2091.60 & 2151.01 & 3103.58 & 3199.14 & 6173.51 \\  
(with internal links)& ($\pm3.58$) & ($\pm$7.84) & ($\pm70.64$) & ($\pm110.08$) & ($\pm70.87$) & ($\pm75.27$) \\
\midrule
SUMO & \multirow{2}{*}{1218.53} & \multirow{2}{*}{1987.12} & \multirow{2}{*}{2046.63} & \multirow{2}{*}{2973.50} & \multirow{2}{*}{3020.16} & \multirow{2}{*}{6083.37} \\
(no internal links) & & & & &\\
\bottomrule
\end{tabular}
\caption{Comparison results of efficiency experiments between two setups of SUMO.}
\label{tab:sumo}
\vspace{-0.3cm}
\end{table*}

\paragraph{Baselines Setup} In our experiment, we use the default setting of SUMO (TRACI) and Cityflow. Note that our simulator is a microscopic one. Therefore, we select two open source microscopic traffic simulators as our baselines. The routing model is not activated for all three simulators so it is not relevant if routing parallelization is used. In the experiment presented in our paper, internal links between intersections are used. 

Considering that CBEngine simplifies links between intersections, a question may be raised whether the internal link has a decisively negative impact on the performance of the baseline. To answer this question, we conduct new efficiency experiments on SUMO with no internal links and the same setting otherwise. The comparison result is demonstrated in Table~\ref{tab:sumo}.

According to the result, removing internal links for SUMO does improve the simulation efficiency. However, SUMO’s efficiency is still not considerable compared with that of CBEngine. This is because CBEngine deploys other optimization (e.g. an optimized parallelization architecture) to improve efficiency.

\paragraph{Datasets} The efficiency experiment is conducted on the traffic data from six cities of different scales: Nanchang, Changchun, Jinan, Shenzhen, Hangzhou, and Shanghai. We obtain the road network data from our dataset of CBData. Traffic flows are generated according to the scale of the road network. Two scalability experiments are conducted on real-world road networks from our dataset with scales close to the selected road network size ($\{10, 10^{2}, 10^{3}, 10^{4}, 10^{4.5}\}$ intersections). Traffic flows are generated according to the scale of the road network. All road networks and traffic flows are available in our code provided on Google Drive.\\
\url{https://drive.google.com/file/d/1NHlIR_0CRlRc87SOACBJxt764Ob-z6ca/view?usp=sharing}\\
We also provide a reproducing instruction to help reproduce our experimental results.\\
\url{https://github.com/CityBrainLab/CityBrainLab/blob/main/CBEngine/Reproducing_Instruction.md}

\paragraph{Computing Resources}
All the experiments are conducted on a Ubuntu20.04 system with a 40-core CPU and 128GB RAM. 

\paragraph{Hyperparameters}
The number of threads is chosen as 20 to stay consistent with the number of used cores. Note that, using fewer or more threads lead to worse efficiency for both Cityflow and CBEngine.

\paragraph{Error Bars}
Error bars of the experiment are shown in Table 1 and Figure 3.

\subsection{Learning from Real Traffic Data (Section 3.2)}

In these experiments, we aim to provide demonstrations to show the possibility to use real-world traffic data to optimize the traffic simulator. Hence, there might be further room for improvement if the parameters are tuned carefully.

\subsubsection{Learning to Simulate Driving}

\paragraph{Datasets} The road network of Shenzhen is obtained from our dataset. We obtain the traffic flow (as the input) and the observation of speed (as the ground truth) from the GPS trajectory data of cars in Shenzhen, China for one day. The data covers 123,481 trajectories and comes from personal data providers with consensus.

\paragraph{Optimization Details}
 We use OpenBox as a toolkit to search for parameters. The code can be found at \url{https://github.com/PKU-DAIR/open-box.git} under the MIT license. The start-up hyperparameters are as follows. For the maximum acceleration and deceleration, we set the default value (value where the search starts) as $2.0m/s^2$ and $5.0m/s^2$, respectively. For the speed limit, we set the default value as $11.1m/s$. The number of rounds is set as 20. We use the surrogate type \textit{auto} and optimizer type \textit{auto}. 

\subsubsection{Learning to Simulate Routing}

\paragraph{Datasets} We demonstrate how to learn the routing module on trajectories data of Shenzhen, China for one day. We collect the trajectories with the origin of $\{$Latitude: $22.5405^{\circ}N$, Longitude: $113.967^{\circ}E\}$ and the destination of $\{$Latitude: $22.6164^{\circ}N$, Longitude: $113.853^{\circ}E\}$. The origin is the bus station of the Window of the World, a famous scenic spot in Shenzhen. The destination is a bus station on the highway from Shenzhen to Guangzhou. This origin-destination pair aggregates the most number of different routes in our dataset, while routes of other origin-destination pairs are quite unified. The total number of different routes is 22 and that of trajectories is 118. 

\paragraph{Optimization Details}

We use an RNN-based model as the trajectory generator. The code can be found at \url{https://github.com/benchoi93/TrajGAIL.git} under MIT license. We follow the setting of hyperparameters in the original paper \cite{choi2021trajgail} except for the iteration number since our trajectories are complicated for the generator to learn from. We set the iteration number as 100.

\subsection{Traffic Signal Control and Congestion Pricing (Section 4.2 and 4.3)}

The goal of these two experiments is to provide possible benchmarks for algorithms. Here, we use several typical algorithms to validate these scenarios. Providing comprehensive baseline methods comparison is out of the scope here. Hence, people are welcome to provide more advanced algorithms for these scenarios.

\paragraph{Datasets} We use two real-world datasets to validate CBScenario: Hangzhou and Manhattan. Both datasets are transformed from the traffic data at \url{https://traffic-signal-control.github.io/#open-datasets}, which serves as a widely used benchmark for traffic signal control. We use part of CBData to transform them to the format suitble for CBEngine. The goal of the experiment is to validate the plausibility of our traffic simulation. Therefore, we refer to the widely used benchmark rather than picking up novel cases not being evaluated yet.

\begin{table}[htbp]
\setlength{\tabcolsep}{1mm}{
\begin{tabular}{c|c|c|c}
\toprule
Memory size & Value updating interval & $\epsilon$ & $\gamma$   \\ 
\midrule
5000 & 1 & 0.9 & 0.95 \\
\midrule
\midrule
Learning rate & Target updating interval & $\epsilon_{min}$ & Decay of $\epsilon_{min}$   \\ 
\midrule
0.005 & 20 & 0.2 & 0.995 \\
\bottomrule
\end{tabular}
}
\caption{Hyperparameters of DQN in traffic signal control.}
\label{tab:hyper}
\vspace{-0.3cm}
\end{table}
\begin{table}[htbp]
\setlength{\tabcolsep}{1mm}{
\begin{tabular}{c|c|c}
\toprule
Memory size & Value updating interval & Policy learning rate   \\ 
\midrule
2000 & 1 & 0.001\\
\midrule
\midrule
$\tau$ & Target updating interval & Critic learning rate   \\ 
\midrule
0.125 & 10 & 0.0005\\
\bottomrule
\end{tabular}
}
\caption{Hyperparameters of EBGtoll in congestion pricing.}
\label{tab:hyper2}
\vspace{-0.3cm}
\end{table}

\paragraph{Hyperparameters of Traffic Signal Control}
The traffic in one episode lasts for 1800 seconds.
The DQN method is trained for 50 episodes and the batch size is set as 64. Other hyperparameters are listed in Table~\ref{tab:hyper}.

\paragraph{Hyperparameters of Congestion Pricing}
The traffic in one episode lasts for 10800 seconds. Actions are taken every 540 seconds. We use the fixed time policy as the default traffic signal control policy. For transportation-based methods, we evaluate them in one episode. For the training of EBGtoll, the number of episodes is 200 and the batch size is set as 32. Other hyperparameters are listed in Table~\ref{tab:hyper2}.

\subsection{APIs for Traffic Control}

CBEngine provides various APIs for users to develop novel traffic control policies. The functions supported are listed as follows:
\begin{itemize}
    \item Changing the phase of a traffic signal light
    \item Modifying the speed limit of a road
    \item Changing the route of a vehicle
\end{itemize}

\begin{figure*}[htbp]
\vspace{-0.2cm}

\includegraphics[width=0.86\linewidth]{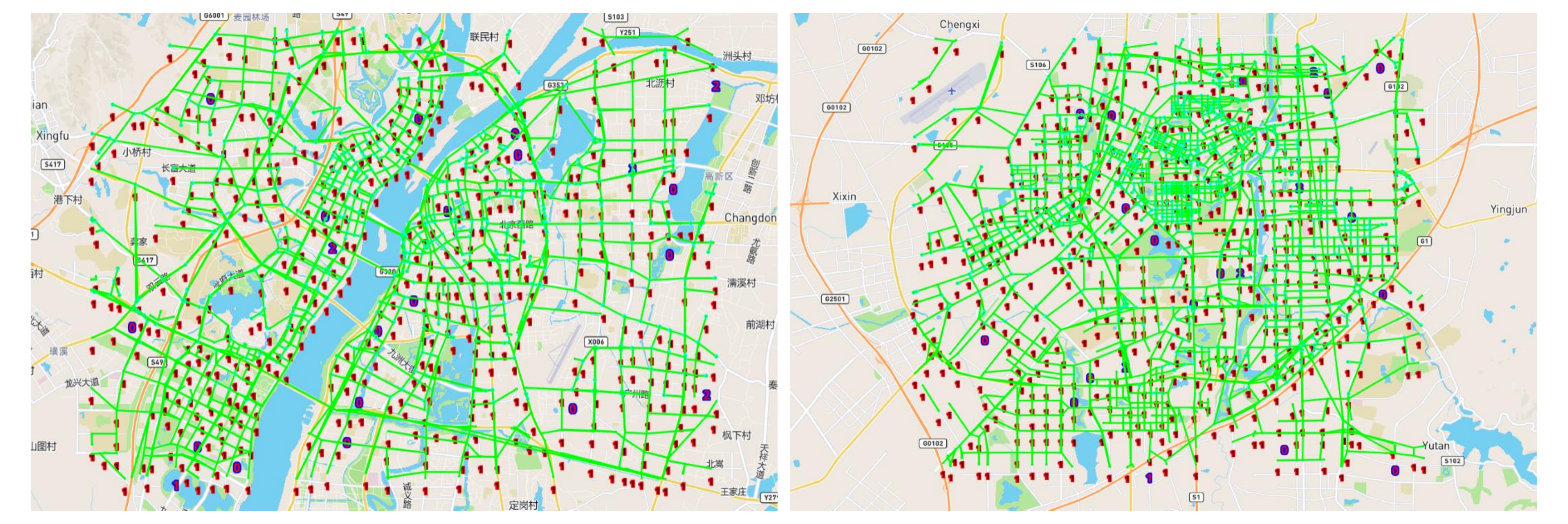}
\includegraphics[width=0.86\linewidth]{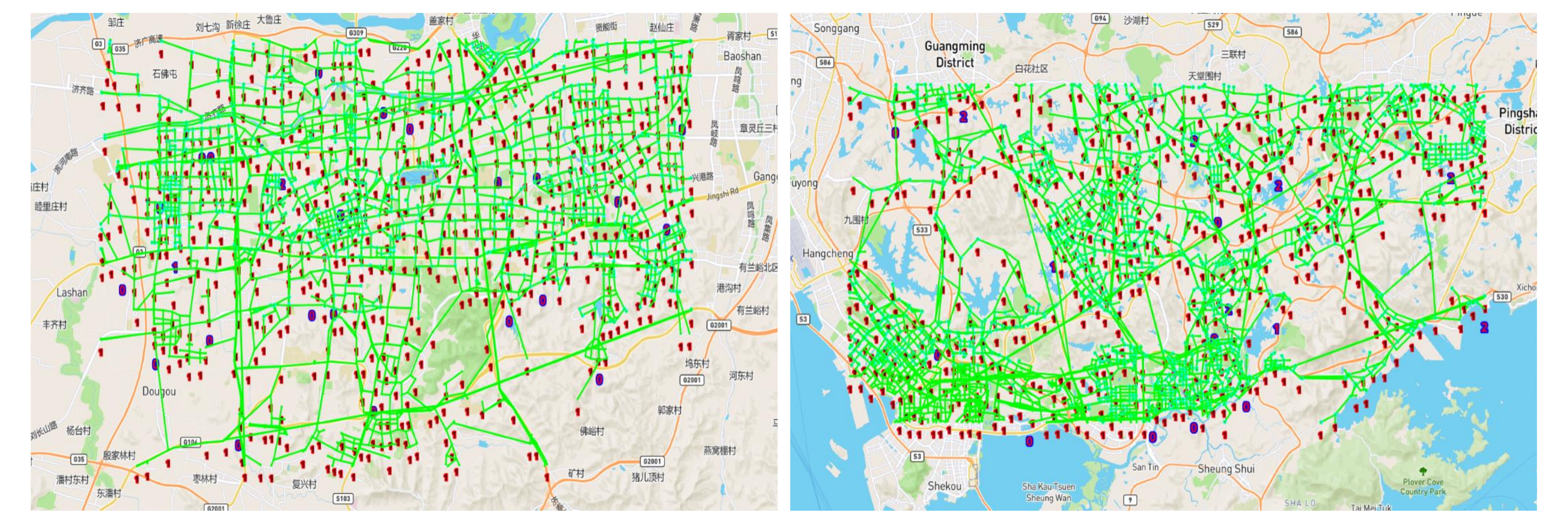}
\includegraphics[width=0.86\linewidth]{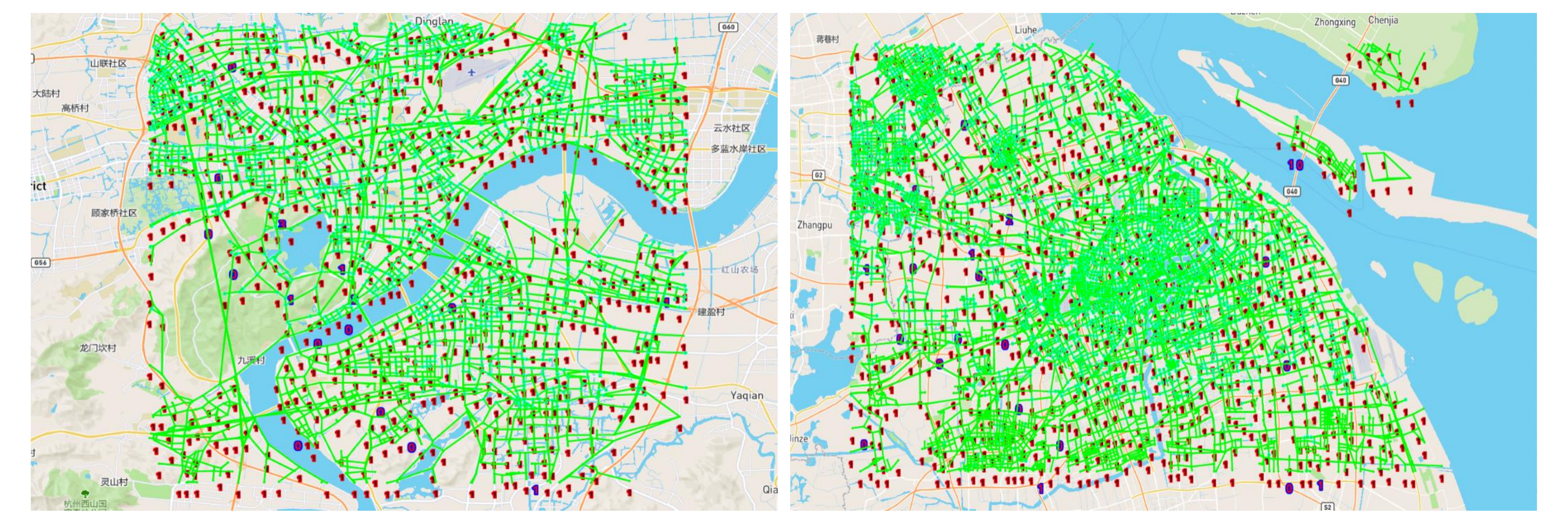}
\caption{Visualization of road networks of six cities used in our experiment: Nanchang (top left), Changchun (top right), JiNan (middle left), Shenzhen (middle right), Hangzhou (bottom left), Shanghai (bottom right). }
	\label{fig:dataset}
\vspace{-0.2cm}
\end{figure*}

\section{City Brain Challenge @ KDD CUP 2021}
\label{appendix:kddcup}
City Brain Challenge @ KDD CUP 2021~\footnote{https://kdd.org/kdd2021/} aimed to explore an efficient decision-making solution for traffic signal control in city-level urban traffic scenarios. The challenge provided an interactive city-level traffic environment with 2,048 intersections, whose data originated from the real urban traffic in Nanchang, China. Players of the challenge were in charge of the urban traffic signals and tried to improve the number of served vehicles and decrease the traffic delay. Each team designed and tuned their plan based on a given traffic flow, while their submitted plan would be then evaluated in another traffic flow. See the documentation~\footnote{https://kddcup2021-citybrainchallenge.readthedocs.io/en/latest/city-brain-challenge.html} for the challenge for more information. 

City Brain Challenge @ KDD CUP 2021 had 1,156 teams of participants. The team \textit{IntelligentLight} from Shanghai Jiao Tong University finally got first place. \textit{IntelligentLight} innovated intelligent traffic signal control algorithms that can improve the traffic efficiency of a city by at least 31\%.

The original version of CBLab supported the interactive city-level traffic environment in the City Brain Challenge @ KDD CUP 2021. Specifically, we built the basic interactive traffic environment on the early version of the simulator CBEngine. After the City Brain Challenge, we introduced more designs to improve the efficiency and scalability of CBEngine, developed the data tool CBData, and expanded the basic interactive environment into CBScenario.

\section{Details of Road Network Dataset}
\label{appendix:dataset}
The road network dataset in CBData includes raw road networks of 100 main cities around the world. The list and the range of these road networks are given on GitHub:\\
\url{https://github.com/CityBrainLab/CityBrainLab/blob/main/CBData/citylist.csv}.
We obtain the data from OpenStreetMap, an open-source map database. Note that the bounding boxes of these road networks have to be a rectangle thus not strictly consistent with the boundary of the city. These road networks can be transformed into road network inputs for traffic simulation with our pipeline in CBData. Here, we visualize six road networks used in our experiment for example to give a scratch of our road network data in Figure~\ref{fig:dataset}.

\end{document}